\documentclass[prx,twocolumn]{revtex4-2}
\usepackage{graphicx}
\usepackage{amsmath}
\usepackage{natbib}
\usepackage{epstopdf}
\usepackage{xcolor}

\newcommand{\n}{\nonumber}
\newcommand{\bn}{\begin{eqnarray}}
\newcommand{\en}{\end{eqnarray}}
\newcommand{\eml}{\end{multline}}
\newcommand{\bml}{\begin{multline}}
\newcommand{\h}{\hspace}

\newcommand{\pt}{\partial_x}
\newcommand{\avgR}{\langle \rho_P \rangle}
\newcommand{\avgL}{\langle \rho_L \rangle}
\newcommand{\avg}{\langle \rho \rangle}
\begin{document}

\title {Quantum Scattering States in a Nonlinear Coherent Medium}

 \author{Allison Brattley$^1$, Hongyi Huang$^2$ and Kunal K. Das$^{2,3}$}
  \affiliation{$^1$ Department of Physics, Massachusetts Institute of Technology, Cambridge, MA 02139, USA}
 \affiliation{$^2$ Department of Physics and Astronomy, Stony Brook University, New York 11794-3800, USA}
 \affiliation{$^3$ Department of Physical Sciences, Kutztown University of Pennsylvania, Kutztown, Pennsylvania 19530, USA}

\begin{abstract}
We present a comprehensive study of stationary states in a coherent medium with a quadratic or Kerr nonlinearity in the presence of localized potentials in one dimension (1D) for both positive and negative signs of the nonlinear term, as well as for barriers and wells. The description is in terms of the nonlinear Schr\"odinger equation (NLSE) and hence applicable to a variety of systems, including interacting ultracold atoms in the mean field regime and light propagation in optical fibers. We determine the full landscape of solutions, in terms of a potential step and build solutions for rectangular barrier and well potentials. It is shown that all the solutions can be expressed in terms of a Jacobi elliptic function with the inclusion of a complex-valued phase shift. Our solution method relies on the roots of a cubic polynomial associated with a hydrodynamic picture, which provides a simple classification of all the solutions, both bounded and unbounded, while the boundary conditions are intuitively visualized as intersections of phase space curves. We compare solutions for open boundary conditions with those for a barrier potential on a ring, and also show that numerically computed solutions for smooth barriers agree qualitatively with analytical solutions for rectangular barriers. A stability analysis of solutions based on the Bogoliubov equations for fluctuations show that persistent instabilities are localized at sharp boundaries, and are predicated by the relation of the mean density change across the boundary to the value of the derivative of the density at the edge. We examine the scattering of a wavepacket by a barrier potential and show that at any instant the scattered states are well described by the stationary solutions we obtain, indicating applications of our results and methods to nonlinear scattering problems.
\end{abstract}
\maketitle

\section{Introduction}

Scattering by a localized potential is one of the fundamental paradigms of quantum mechanics, defining the dynamics and interactions of many body systems \cite{mahan}. The linear problem that defines scattering of non-interacting particles in one dimension is part of any introduction to quantum physics \cite{liboff}. On the other hand, the nonlinear problem of scattering of interacting particles is a substantially more complex problem and a comprehensive picture is lacking. A close formal analog to the linear problem can be found in the mean field description of the scattering of interacting bosons in terms of a nonlinear Schr\"odinger equation (NLSE), where the effect of inter-particle interactions appears in the form of an added quadratic nonlinear term \cite{RMP-Sringari-1999}.

The nonlinear Schr\"odinger equation arises in multiple contexts, its initial applications in 1D propagation being in the context of self-focussing of light \cite{NLSE-Chiao-Townes}, and thereafter much of the subsequent studies were in the field of nonlinear and fiber optics \cite{Boyd, G.P.Agarwal}. In the last few decades, with the creation of Bose-Einstein condensates (BEC) \cite{Pitaevskii-Stringari,Pethick-Smith}, it took on a revitalized role as the Gross-Pitaevskii equation, which dominated the early theoretical description of BEC and continues to be relevant in the mean field regime that captures much of the stationary and dynamical properties of large condensates \cite{Kevrekidis-Carretero}.

There is a vast literature on the NLSE and its solutions \cite{Khawaja-NLSE,Fibich-NLSE}. However, with some notable exceptions we discuss below, in the context of both optical and matter waves, prior works fall into two categories:  Analytical and numerical stationary solutions obtained assuming a uniform system, without a potential \cite{Zakharov-JETP, A1, A2, B13, B34} or with a periodic lattice \cite{A4, B10, B11, B15, B16}; or time-dependent problems focused on the propagation, dispersion and, occasionally, scattering of localized soliton wavepackets. The latter is largely driven by applications in optical communications with influential theoretical work \cite{B21_1,B21_2, B14,B18} supported by experiments \cite{B20, B26, B27, B35, B32}. The creation of BEC provided a new paradigm for soliton studies with the intrinsic quantum nature and massive character of matter waves driving interest in scattering dynamics of solitons by local potentials, in theory \cite{B3,B4,B5,A10, A11, A12, A13}  as well as in experiments \cite{A8,A9}. Active interest continues with possibilities of probing quantum nonlocality with macroscopic superpositions involving soliton pairs \cite{Kevrekidis-expt-oscillating-solitons, Castin-nonlocal-superposition,Cederbaum-BEC-attractive} and with recent realization of matter wave counterparts of breathers \cite{Olashanii-Hulet-breathers, Olashanii-Hulet-fluctuations} previously studied only in optical systems \cite{Optical-breathers}.

In contrast, there are few studies of the stationary solutions of the NLSE in the presence of a localized potential, analogous to the linear scattering problem; and they are generally limited by different constraining assumptions. The delta potential along with the step potential were examined in \cite{A3}, and the dynamics in the presence of delta potential impurities in \cite{Hakim-NLSE_delta_potential,Pavlov-BEC_delta_potential}; bound states in a square well were examined in \cite{A5}; solutions that neglect nonlinearity outside the potential were studied in \cite{A6}; a perturbative study in the limit of weak nonlinearity was done in \cite{A7}. Although the superposition principle does not apply in the nonlinear problem, an approximate form of  was assumed in the studies in \cite{A6, A7}, but not in some subsequent studies of stationary solutions that specifically examined rectangular barriers for repulsive interactions \cite{A0, A15}. In addition to the limiting assumptions indicated, the studies were also restricted to solutions that were bounded at infinity, with one notable exception \cite{A15} which considered a specific type of bilaterally symmetric unbounded ones in the context BEC flow through a weak link. This, as we show here, leaves out a large class of solutions. Furthermore, certain assumptions were made about the parameters involved that are at best incomplete, as we will describe in the relevant sections.

The purpose of this paper is to provide a comprehensive landscape of analytically obtained solutions in the presence of a localized rectangular potential in one spatial dimension for a system describable by a quadratic nonlinear Schr\"odinger equation,  for both positive and negative interactions and for barriers and wells;  this includes solutions that are bounded as well as unbounded at infinity, the latter allowed within a finite width potential. All previously studied cases can be obtained as subsets or limiting cases of our solutions. Notably, typical descriptions of single solitons in terms of hyperbolic functions are limiting cases of our solutions in terms of Jacobi elliptic functions \cite{Handbook_elliptic}.

Such stationary solutions will provide the basis for describing diverse nonlinear dynamical phenomena, including the scattering of soliton and solitonic trains in optical or cold atomic systems, and the relative motion of a potential through a superfluid \cite{BEC-flow-Atherton, Hulet-BEC-transport}. The latter has garnered much recent interest in the context of persistent currents in BEC in ring configurations \cite{ramanathan,Phillips_Campbell_superfluid_2013} due to sustainable superfluidity.  We also find solutions for a potential localized along the azimuth here for such ring configurations in this paper; elsewhere we have found analogous solutions in the presence of a lattice \cite{Das-Huang,Das-Tekverk-Siebor}.  Using separate, numerical simulations we also show here that our analytical solutions are in qualitative agreement with solutions for smooth barriers with profiles similar to those associated with focussed lasers used in experiments on persistent flow.

A significant outcome this work is to demonstrate that our stationary solutions can have direct utility in the description of scattering of even non-uniform wavepackets, by numerically scattering a wavepacket on a rectangular barrier and mapping out the resulting transmitted and reflected densities in terms of a finite range of analytical solutions. This establishes a novel and more reliable approach to applying analytical solutions to scattering problems in the absence of the superposition principle.

The paper is organized as follows: Section II presents the physical model and defines the general form of the solutions and the roots-based approach we will utilize, with Sec.~III defining the physically imposed constraints on those roots. Section IV defines the crucial impact of boundary conditions at the potential edges. The effects are contrasted with the linear limit of zero nonlinearity in Sec.~V, which helps us understand how the density changes across the potential boundary when the nonlinearity is introduced in Sec.~VI. In Sec.~VII, we show the solutions can have a complex phase shift and describe the significant effects and constraints accompanying such shifts for both positive and negative nonlinearities. Sections VIII and IX determine the allowed solutions for a step potential for positive and negative nonlinearities respectively. These are then applied to barrier and well potentials in Secs. X and XI respectively. In Sec.~XII, we study the limitations on the solutions arising from having a ring structure. Section XIII compares our solutions with those obtained via numerical simulations for a smooth barrier. We conduct a general stability analysis for our solutions in Sec.~XIV based on the Bogoliubov equations \cite{Pitaevskii-Stringari} for fluctuations.  In Sec.~XV, we use our solutions to analyze the scattering of a wavepacket on a localized barrier.  We summarize  our conclusions and future outlook in Sec.~XVI. Two appendices provides some details of our calculations and derivations.

\section{Physical Model}
We consider the nonlinear Schr\"odinger equation:
\begin{equation}
\label{eq1.1-9}
\left[-{\textstyle\frac{1}{2}}\partial_x^2+V+g|\psi|^2\right]\psi=-i\partial_t\psi
\end{equation}
In the context of ultracold atoms which motivates this study, this is a mean field equation for the expectation of the bosonic field operator $\langle\hat{\Psi}\rangle=\psi$ \cite{Pitaevskii-Stringari,Pethick-Smith}. The one dimensional description can be considered an effective picture with transverse degrees of freedom integrated out due to tight confinement \cite{Das-PRA-R-Shift}. Taking that to be cylindrical and harmonic, the trap frequency $\omega$ can be taken to set our units $l=\sqrt{\hbar/(m\omega)},
\epsilon=\hbar\omega$ and $\tau=\omega^{-1}$, and the  effective 1D
interaction is $g=2a$ defined by the scattering length $a$ . Assuming infinite extent typical of scattering problems, $|\psi|^2$ defines the number density. The stationary solutions $\varphi(x)=\psi(x,t)e^{i\mu t}$ satisfy the time-independent version of Eq.~(\ref{eq1.1-9}) with $i\partial_t\rightarrow \mu$ where the eigenvalues $\mu$  define the chemical potential.

\begin{figure*}[t]
\centering
\includegraphics[width=\textwidth]{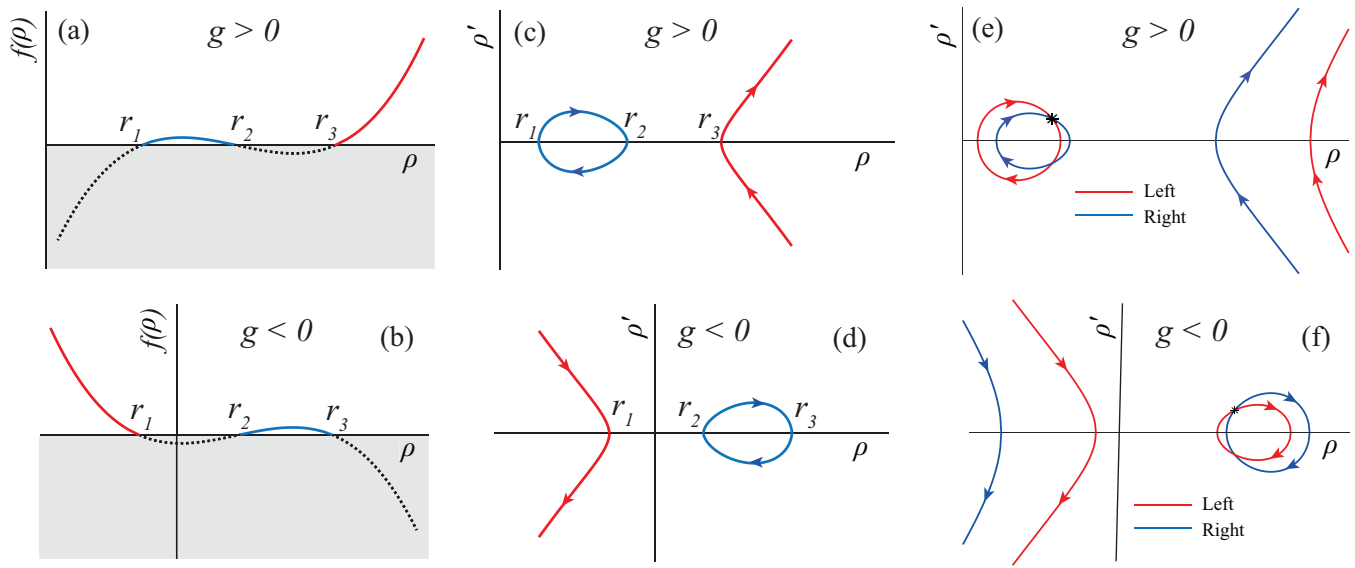}
\caption{Features of phase space curves are illustrated with upper panels for positive nonlinearity $g>0$ and lower panels for negative nonlinearity $g<0$. (a,b) Schematics of  the cubic function $f$ that sets the density variation, for the case of all three real roots $\{r_i\}$; the dotted parts lie in the shaded forbidden region. (c,d) Corresponding phase space plot of $\rho'=\pm\sqrt{f}$ versus $\rho$, showing the wing-loop structure; the orientation of the wing depends on the sign of the nonlinearity $g$. (e,f) Intersection of curves from different potential regions (labelled `left' and `right') determine the matching of solutions at their boundary. }
\label{Fig1_functionshape}
\end{figure*}

The behavior of the physical observables can be understood best by writing the mean field stationary state in the polar amplitude-angle form, referred to as the hydrodynamic picture, $\varphi( x)=\sqrt{\rho( x)}e^{i\phi( x)}$, leading to an equation for the density $\rho$
\begin{equation}
\label{eq1.2-2}
{\textstyle\frac{1}{8}}(\pt\rho)^2-{\textstyle\frac{1}{4}}\rho\pt^2\rho+{\textstyle\frac{1}{2}}
\alpha^2+V\rho^2+g\rho^3-\mu\rho^2=0
\end{equation}
and a phase equation that provides an integral of motion,
\bn\label{eq1.2-1}\textstyle
\rho\pt\phi-\Omega\rho&=&\alpha\\ \Delta \phi ( x)=\phi ( x)-\phi (0) &=&\Omega x+\int_{0}^{ x}\frac{\alpha}{\rho( x')}d x'.
\n\en
This sets the current density $J=\alpha$, the superfluid velocity $v=\alpha/\rho( x)$ and angular momentum per particle $L=\hbar\Omega+2\pi\hbar\alpha$.  We include an optional $\Omega$ to allow for rotation in the case of a finite sized ring with periodic boundary condition that we also consider \cite{Das-Huang}.

A first integration of Eq.~(\ref{eq1.2-2}) yields
\bn
\label{eq.hd.3}
\pt\rho&=&\pm\sqrt{f\left(\rho\right)}\n\\
f\left(\rho\right)&=&4g\rho^3-8(\mu-V)\rho^2+8\beta\rho-4\alpha^2
\en
with integral of motion $\beta$ which can be written in terms of the integrated expression
\bn \beta={\textstyle\frac{1}{8\rho}}(\pt\rho)^2+{\textstyle\frac{1}{2\rho}}
\alpha^2-{\textstyle\frac{1}{2}}g\rho^2+(\mu-V)\rho.\label{eq.phase}\en
For a \emph{constant} potential $V$, the density solution is a Jacobi elliptic function
\bn
\label{eq.jacobidens}
\rho(x)&=& r_1+(r_2-r_1){\rm sn} ^2(\textstyle{\sqrt{g(r_3-r_1)}}\ x +x_0,m)
\en
expressed here in terms of the roots of the cubic polynomial  $f=4g(\rho-r_1)(\rho-r_2)(\rho-r_3)$.  Those roots determine most of the relevant parameters
\bn\label{parameters}\textstyle
m&=&\frac{r_2-r_1}{r_3-r_1},\h{5mm} \mu-V=\frac{g}{2}(r_1+r_2+r_3)\n\\
\alpha^2&=& gr_1 r_2 r_3,\h{5mm}
\beta=\frac{g}{2}(r_1 r_2+r_1 r_3+r_2 r_3).
\en

The additional parameter $x_0$ represents translations necessary to match the boundary conditions at the interface of different potential regions. The general form of the solution is given by Eq.~(\ref{eq.jacobidens}) for the density and Eq.~(\ref{eq1.2-1}) for the phase. Solutions that satisfy the physical constraints and boundary conditions are determined by the cubic function $f(\rho)$ in Eq.~(\ref{eq.hd.3}), illustrated in Fig.~\ref{Fig1_functionshape}(a,b) for positive and negative nonlinearity respectively. Since it needs to be positive definite, the dotted segments are forbidden. On plotting $\rho'=\pm\sqrt{f}$, the positive and negative branches match up smoothly along $\rho=0$ to create the shapes shown in the adjacent panels Fig.~\ref{Fig1_functionshape}(c,d).

These phase space plots of $\rho'$ versus $\rho$ plots have a characteristic structure that will guide much of our analysis.  They can typically have a closed loop and an open boomerang shape which we will refer to as the `\emph{loop}' and `\emph{wing}', respectively. Alternately, when $f(\rho)$ intersects the axis at only one point, the loop and the wing merge to create a \emph{conjoined} profile. There are concrete physical implications of these segments that we will discuss at length, but the most basic one is that the loop signifies oscillating solutions, whereas the wing or the conjoined parts correspond to solutions that do not oscillate and can be unbounded.

If the derivative $\rho'>0$, then the density $\rho$ has to increase, and if $\rho'<0$, then $\rho$ has to decrease. This means that the variation in density can follow the loop only in the \emph{clockwise} orientation. This is true for both positive and negative nonlinearities. However, as shown in Fig.~\ref{Fig1_functionshape}(c,d), the wing opens in opposite orientations for  positive and negative nonlinearities. This means that for $g>0$, the density associated with the wing varies such that it approaches the horizontal axis ($\rho'=0$) from below and moves away from it above; while for $g<0$, the behavior is opposite. This will have important implications when there are intersections of phase space curves from different potential regimes at their boundary, as sketched for loop segments in Fig.~\ref{Fig1_functionshape}(e,f).

\begin{figure}[t]
\centering
\includegraphics[width=\columnwidth]{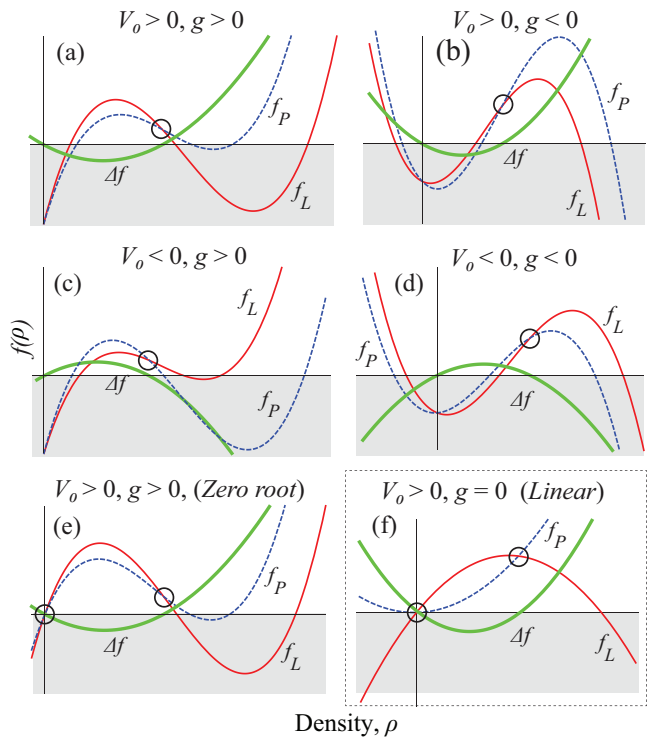}
\caption{(Color online) (a-e) The cubic function $f(\rho)$ is plotted for both side of a potential boundary, $f_L$ with $V_0=0$ and $f_P$ for $V_0\neq 0$, along with their difference $\Delta f$, for various combinations of repulsive $g>0$ or attractive $g<0$ nonlinearity and potential barrier $V_0>0$ or well$V_0<0$. The $f$ from the two regimes generally intersect at two points, one being always at $\rho=0$. The physically relevant intersections are marked by circles, shaded regions being forbidden. (f) In the linear case $V_0=0$, the function $f$ is quadratic, here a special case is shown where both intersections are physical.}
\label{Fig2_fp_minus_fl}
\end{figure}

\section{Constraints on Roots}\label{SecIII:Constraints}

There are several factors that restrict the allowed space of solutions. We start with some general considerations. The ordering $r_1<r_2<r_3$ will always be assumed when the roots are real. The definition of the parameter $\alpha^2=gr_1r_2r_3$ and the requirement that $\alpha$ has to be real to be physically relevant both constrain the allowed roots.

For positive nonlinearity, when $g>0$, there are three cases: (i) Real roots have to be either all non-negative $r_1,r_2,r_3\geq 0$ or (ii) one can be non-negative and the other two negative or zero, $r_3\geq0$ and $r_1,r_2\leq 0$; (iii) Alternatively, there can be one non-negative root and two complex conjugate roots, $r_1\geq0$ and $r_2=r_3^*$. The last case has been often disregarded in prior studies \cite{A7,A0}, since the existence of complex conjugate roots implies both a complex elliptic modulus and a complex coefficient of the squared Jacobi elliptic function in Eq.~(\ref{eq.jacobidens}); but we will show here that they still provide physical solutions. Together, these three combinations define the space of the possible types of solutions for the density in a given region. Notably for $g>0$, since at least one root has to be positive, the wing or conjoined structures of the phase space curve must always intersect the $\rho$-axis at $\rho\geq 0$. Furthermore, because they open on the right  as shown in Fig.~\ref{Fig1_functionshape}(c), they must lie in their entirety in the $\rho>0$ regime; the loop part has no such restrictions.

For negative nonlinearity, $g<0$, at least one root has to be less than or equal to zero, but two roots cannot be negative and three negative roots have no physical meaning. Complex roots are not allowed since a conjugate pair would have a positive norm. This means that the only possibility is for all the roots to be real with $r_1\leq 0$, and $r_2,r_3 \geq 0$; and $r_1$ marks the intersection of the wing with the $\rho=0$ axis. Since, the wing opens on the left, shown in Fig.~\ref{Fig1_functionshape}(d), this also means for $g<0$, the entirety of the wing lies in the $\rho\leq0$ regime.

Notably, if one of the three roots is zero in \emph{any} region, with or without potential, then \emph{all} of the regions of the system will necessarily have a zero root.  This simply means that $\alpha$, which is a measure of the current, needs to be conserved across the system.

\section{Boundary Conditions}\label{SecIV:Boundary}
We will consider both step potentials and rectangular potentials. In order to keep the notation consistent, we will label the region of the non-vanishing potential with subscript $P$ and, for a rectangular barrier or well, the left and the right of it with subscripts $L$ and $R$, while for a step potential we will only have $L$ and $P$ regions. The basic element of our system is a potential step, and matching the boundary conditions at the edge of a step determines the complete solutions. Without loss of generality, such a potential can be described by  $V(x)=V_0\Theta(x)$ where $\Theta(x)$ is the Heaviside step function and $V_0$ can be either positive or negative.

The chemical potential and current density are conserved throughout the system, across segments with and without a potential.  In contrast, the parameter $\beta$ is fixed only within each region, but changes as the potential changes from one region to the next.  Since $V=0$ for $f_L$ and $V\neq 0$ for $f_P$, from Eq.~(\ref{eq.hd.3}) it follows
\bn \label{eq_f}  \beta_P &=& \beta_L-V_0\rho_0\n\\
\Delta f&=&f_P(\rho)-f_L(\rho) = 8V_0\rho(\rho-\rho_0)\en
with the values in the different regions set by the density at the boundary $\rho_L(x=0)=\rho_P(x=0)=\rho_0$. The edge of the potential is taken as our co-ordinate origin. At the boundary, clearly $f_L(\rho_0)=f_P(\rho_0)=f_0$, which serves as the definition of $f_0$ as the common value at the boundary.

Equation~(\ref{eq_f}) implies that $f_P=f_L$ at only two points: at $\rho=0$ and at $\rho=\rho_0$. Therefore the functions $f_P$ and $f_L$ in the regimes with and without the potential can intersect only at those two points, which will therefore set the boundary conditions.  Figure~\ref{Fig2_fp_minus_fl} illustrates this for different scenarios we will consider. The difference $\Delta f=f_p-f_L$ is a parabolic function of the density $\rho$, represented by a thick green line; it is concave upwards for $V_0>0$ in panels Fig.~\ref{Fig2_fp_minus_fl}(a,b,e,f) and concave downwards for $V_0<0$ in Fig.~\ref{Fig2_fp_minus_fl}(c,d). Furthermore, Fig.~\ref{Fig2_fp_minus_fl}(a,c,e) have positive nonlinearity $g>0$ whereas Fig.~\ref{Fig2_fp_minus_fl}(b,d) have negative nonlinearity $g<0$. Since $f(\rho)\geq 0$, the shaded region below the $\rho$ axis is not allowed, therefore, in most cases, only one intersection is physical.  The physically relevant intersections between $f_P$ and $f_L$ are marked by black circles in Fig.~\ref{Fig2_fp_minus_fl}. A second intersection can be seen in all the panels and always occurs at $\rho=0$, but it is unmarked if it occurs in the nonphysical shaded region.

The intersections in $f$ manifest as intersections in `\emph{loop-wing}/\emph{conjoined}' structures appearing in the square root $\rho'=\pm\sqrt{f_0}$ as illustrated earlier in the phase space plots in Fig.~\ref{Fig1_functionshape}(e,f). We can conclude that the phase space curves for the two regimes can intersect only at a maximum of three points, with two of them being the positive and negative roots of $\pm\sqrt{f_0}$, appearing symmetrically above and below the $\rho$ axis as shown Fig.~\ref{Fig1_functionshape}(e,f). A possible third intersection can occur at $\rho=0$, not shown in that figure, but corresponding to allowed intersections such as in Fig.~\ref{Fig2_fp_minus_fl}(e).

When there are three intersections, the density at the boundary can never be $\rho_0=0$ because then the RHS of Eq.~(\ref{eq_f}) becomes $8V\rho^2$, and that corresponds to a single point of intersection of the two curves only at $\rho=0$.  So, although it is possible that the density can vanish at the boundary, $\rho_0=0$, that can only occur if that is the sole intersection in the phase space plots.

Clearly, there can be intersections between the loop, wing or conjoined structure of $\rho'$ of one regime with any one of those from the other regimes, leading to different pairings of solutions across the potential boundary. But, as our analysis in the following sections will demonstrate, not all combinations are physically possible.

\section{Linear Limit}

It is interesting to consider a linear system for comparison. With $g=0$, $f(\rho)$ is a quadratic with a parabolic shape, and the two roots of the equation yield
\bn \alpha^2 = 2(\mu-V_0) r_1 r_2,\h{3mm}
\beta =(\mu-V_0)(r_1+r_2), \label{linear_limit}\en
and $\mu$ is a free parameter and identical to the total energy.  When $\mu>V_0$, the parabola opens downward, like $f_L$ in Fig.~\ref{Fig2_fp_minus_fl}(f), and when $\mu<V_0$ it opens upwards as assumed for $f_P$ in that same figure.

For $\mu<V_0$ physically relevant $\alpha$ requires either (i) $r_1<0,r_2>0$ else (ii) $r_1=r_2=0$. Complex conjugate pair of roots are not possible, their product being always positive, ruling out cases with a minimum at $f_P>0$ . Case (ii) for $f_P$ is shown in Fig.~\ref{Fig2_fp_minus_fl}(f)  and corresponds to solutions that decay in the region of the potential as $e^{-\kappa x}$ with $\kappa=V_0-\mu$ . The more general case (i) where the minimum is at $f_P<0$ corresponds to linear combination of $Ae^{-\kappa x}+Be^{\kappa x}$ within the potential.  These are ruled out for a step potential since they blow up as $x\rightarrow \pm\infty$, but are valid solutions for a finite width potential, a trend we will see often for the nonlinear problem.

When $\mu>V_0$ even in the potential region, both parabolas are downward-facing, and we have oscillatory solutions in both regions, with at least one physical intersection of $f_p$ and $f_L$ at some $\rho_0>0$ similarly to the nonlinear case. If there is a second intersection at $\rho=0$, one of the roots $r_1=0$ and so the current vanishes with $\alpha=0$ and the oscillations have nodes. When $r_1=r_2\neq 0$ the solutions are plane waves.  Plane wave solutions for non-zero degenerate roots are impossible with an upwards parabola since they need to satisfy $r_1r_2\leq 0$.

We now anticipate the considerations of the next section by first illustrating them with the linear case. Leaving aside the  case of exponential decay, the solutions are sinusoidal or plane waves, so the average density is given by $\avg=\frac{r_1+r_2}{2}$, with plane waves corresponding to degenerate roots. Using the expression for $\beta$ above in Eq.~(\ref{linear_limit}) in conjunction with the variation of the $\beta$ across a potential step in Eq.~(\ref{eq_f}) we obtain a relation between the mean densities in the regions with ($P$) and without ($L$) the potential
\bn 2\mu\left(\avgR-\avgL\right) = V_0\left(2\avgR-\rho_0\right) \en
It is obvious that the density at the boundary $\rho_0$ has to have a value between the maximum and minimum values of the density in either region, which means that for a plane wave or oscillating solutions $2\avgR\geq\rho_0$, so for $V_0>0$, $RHS\geq0$ in the above relation. Since $\mu\ge 0$, being the kinetic energy, it follows that $\avgR\geq\avgL$. The mean value of the density therefore increases in the region of higher potential.  This is consistent with current conservation: In a region of higher potential, the net velocity is lower, therefore the mean density must be higher.

For solutions that decay within the potential, the average density would approach zero in the limit of infinite extent of a step. Even for a finite width barrier, when the average density does not vanish,  current would vanish within the potential since both roots are zero. This is consistent with current conservation because, as Fig.~\ref{Fig2_fp_minus_fl}(f) shows, the function $f_L$ outside the potential also has a zero root, so that the density oscillations have a node implying vanishing current.

\begin{figure*}[t]
\centering
\includegraphics[width=\textwidth]{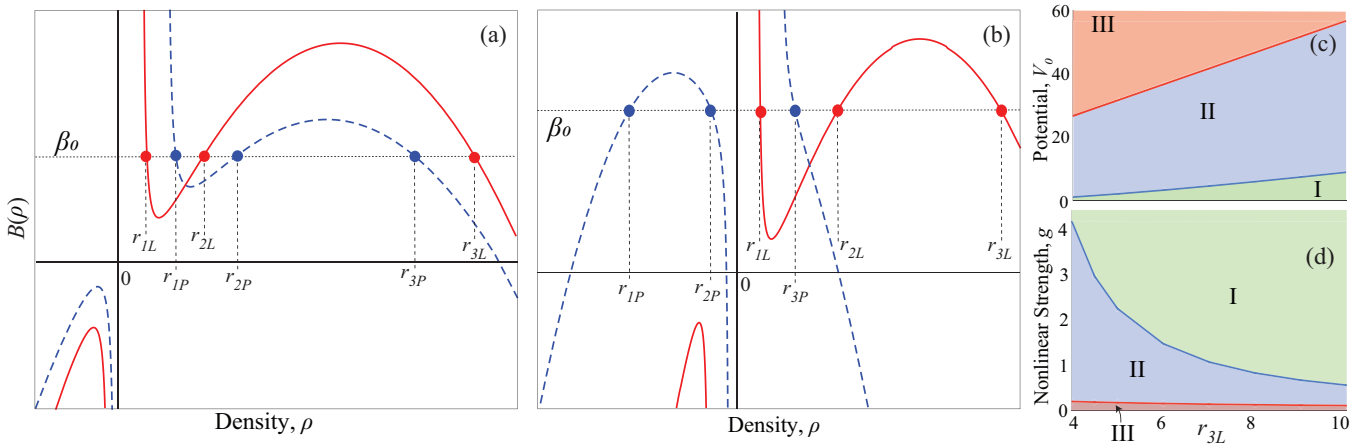}
\caption{(Color online) The function $B(\rho)$ is plotted with solid red line for the left region $V_0=0$ and blue dashed line for the region of the potential $V_0>0$. Any horizontal slice set by a value $B=\beta_0$ (thin dotted line) here mark the roots at points of intersections and show their migration:  Transition of an oscillating to another oscillating function is shown in (a) and to a wing solution in (b) where the negative part of the function rises to meet the $\beta_0$ line  as the potential $V_0$ increases. (c) Fixing parameters on the left for $V_0=0$, except for one root $r_{3L}$, as the potential $V_0>0$ increases the density transitions from being oscillatory (\emph{I, green}) to non-oscillatory first on a conjoined phase space curve (\emph{II, blue}) then on the wing of a loop-wing structure (\emph{III, red}).  (d) The transition of the solutions when $g$ is varied at fixed $V_0$, with same legend as in (c).}
\label{Fig3_roots_migration}
\end{figure*}

\section{Density Change across Boundary}\label{SecVI:Density_Change}

Given a solution on one side of a step boundary, we would like to determine the solution on the other side. Since the solutions are determined by the roots of $f(\rho)$, we need to understand how the roots migrate across a potential boundary, specifically the relation of the roots $\{r_{1L},r_{1L},r_{1L}\}$ on the left with $V_0=0$ with the roots $\{r_{1P},r_{1P},r_{1P}\}$ in the region of the potential $V_0\neq 0$.

We now track the migration of the roots across the boundary for the nonlinear problem to predict the change of the mean density. Unlike in the linear case in the previous section, now $\mu$ as well as $\alpha$ and $\beta$ are fixed by the roots as shown in Eq.~(\ref{parameters}).  Using those expressions, we rewrite the relation in Eq.~(\ref{eq_f}) for  $\beta$ as
\bn &&\h{-5mm}\frac{\alpha^2}{gr_{nL}}+r_{nL}\left(\frac{2\mu}{g}-r_{nL}\right)\\&& = \frac{\alpha^2}{gr_{nP}}+r_{nP}\left(\frac{2\mu}{g}-r_{nP}\right)+\frac{2V_0}{g}(\rho_0-r_{nP}). \label{beta_equation}\n\en
Due to symmetry with respect to the exchange of the roots in the expressions for $\alpha,\beta$ and $\mu$, here $r_n$ represents any one of the three roots. Thus, the boundary condition above can be expressed in terms of a function
\bn B(\rho; V_0) = \frac{\alpha^2}{g\rho}+\rho\left(\frac{2\mu}{g}-\rho\right)+\frac{2V_0}{g}(\rho_0-\rho) \en
as $B(r_{nL};V_0=0)=B(r_{nP};V_0\neq0)=\beta_0$, where we can interpret $\beta_0$ as the constant of integration for $V_0=0$.  Considered as two separate equations, the roots in any region of constant potential is determined by $B(\rho)=\beta_0$.  This is illustrated in Fig.~\ref{Fig3_roots_migration} (a,b), where the function $B$ for each region is plotted. The intersections of the horizontal dotted line, marking a specific value of $\beta_0$,  with each curve determines the roots in the corresponding potential regime.  Note the two curves mutually intersect at $\rho_0$, as should be apparent from Eq.~(\ref{beta_equation}).

As the function $B(\rho;V_0)$ shifts with changes in the potential, like in Fig.~\ref{Fig3_roots_migration}(a,b), the value of the roots, as set by the intersections of $B(\rho;V_0)$ curves with the line of constant $\beta_0$, will shift as well. For oscillating functions, we can easily predict which way the roots will shift, because $r_{1P}\leq\rho_0$, $r_{1L}\leq r_{1P}$, and likewise for $r_{2P}$ and $r_{3P}$, we must have $r_{2L} \leq r_{2P}$ and $r_{3L} \geq r_{3P}$, as shown in Fig.~\ref{Fig3_roots_migration}(a). At a critical $V_0$ there is only one intersection with the $\beta_0$ line, corresponding to one real root and two complex roots. Increasing $V_0$ further leads to negative density regimes, where the curves $B(\rho<0; V_0)$ migrate upwards to intersect the $\beta_0$ line leading to two negative roots and one positive root, as shown in Fig.~\ref{Fig3_roots_migration}(b).

Consider oscillating solutions on both sides such as in Fig.~\ref{Fig3_roots_migration}(a). The nonlinear functions are no longer sinusoidal, so we can no longer assume the mean density to be the average of the roots. However, for $g>0$, the Jacobi elliptic function that defines the density Eq.~(\ref{eq.jacobidens}), becomes more `flat-topped' (see Fig.~\ref{Fig9_barrier}(b) for an example) and the mean value increases with the elliptic parameter $m$, so the linear limit $m=0$ when the density is sinusoidal marks the minimum $ \avg\geq\avg_{min} = \frac{r_1+r_2}{2} $. As the migration of the roots dictate, $m=\frac{r_2-r_1}{r_3-r_1}$ will be larger within the potential, so $\avgR-\avgR_{min}\geq \avgL-\avgL_{min}$ and $\avgR_{min}>\avgL_{min}$. Together they imply that the mean value increases under the influence of a positive potential step, $\avgR>\avgL$. The opposite trend is seen for a negative step, $V_0<0$; $r_1$ and $r_2$ decrease while $r_3$ increases within a potential well, $\avgR<\avgL$.

If we fix the parameters on the left and increase the potential $V_0$ on the right, the density within the potential step will transition from being oscillatory on the loop part of phase space curve, to being non-oscillatory first on a conjoined curve then on the wing of a loop-wing structure. Those phases are marked in Fig~\ref{Fig3_roots_migration}(c) as a function of $V_0$ and $r_{3L}$ for fixed $r_{1L},r_{2L}$ .  As $V_0$ increases, $r_2,r_3$ eventually merge and become complex conjugates, creating the conjoined structure; further increase causes a new loop to emerge on the left as in Fig.~\ref{Fig3_roots_migration}(b), so the sole real root $r_{1P}\rightarrow r_{3P}$ now lies on a wing. The opposite trend is seen in Fig~\ref{Fig3_roots_migration}(d) for the solutions on increasing nonlinear strength $g$ keeping $V_0$ fixed.

\section{Effects of a Complex Phase Shift}\label{SecVII:PhaseShift}

\begin{figure}[t]
\centering
\includegraphics[width=\columnwidth]{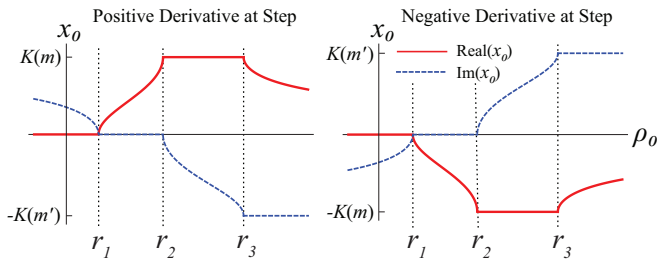}
\caption{(Color online) The real and imaginary values of the phase shift $x_0$ at edge of the potential step, as a function of $\rho_0$ the density at the edge. The real and imaginary parts never vary simultaneously over any density range. As seen, they alternate: one varies while the other remains constant, at either zero or $\pm K(m)$ for ${\rm Re} \{x_0\}$, and zero or $\pm K(m')$ for ${\rm Im} \{x_0\}$. The two panels also show that the sign of $x_0$ depends on the derivative $\rho'(\rho_0)$ at the step boundary.}
\label{Fig8_x0}
\end{figure}

We now turn to a parameter that appears almost arbitrary and is often treated as such, but which, as we now show, actually plays a significant role in defining the solutions. This is the phase shift $x_0$ in Eq.~(\ref{eq.jacobidens}), which sets the density at the origin,
\bn
\label{eq.jacobidens1}
\rho_0&=& r_1+(r_2-r_1){\rm sn} ^2(x_0,m)
\en
Crucially, $x_0$ can in general be complex-valued, a fact generally overlooked, with $x_0$ being tacitly assumed to be real for a solution of this structure \cite{Khawaja-NLSE}. We plot real and imaginary parts of $x_0$ vs $\rho_0$ in Fig.~\ref{Fig8_x0}.  The Jacobi elliptic function ${\rm sn}(u,m)$ is double periodic \cite{Handbook_elliptic} and the density depends on its square. Thus, the values of $x_0$ are plotted modulo those periods, $\mod({\rm Re}\{x_0\},K(m))$ and $\mod({\rm Im}\{x_0\},K(m'))$, where $m'=1-m$ and $K(m)$ is an elliptic integral of the first kind and defines the periodicity of the elliptic functions.  The structure of the function is revealing, establishing that there are indeed strong constraints on $x_0$, contrary to what has been stated in previous papers on this same topic \cite{A0}. The value of $\rho'(x=0)$ simply changes sign of $x_0$ as shown, so our discussion will be in the context $\rho'>0$ in Fig.~\ref{Fig8_x0}(a).

The values of $x_0$ are constrained by the necessity of having real $\rho_0$ value. Specifically, we notice the real and imaginary parts each varies alternately while the other remains constant, as we illustrate with Fig.~\ref{Fig8_x0}(a): When $\rho_0\leq r_1$,  ${\rm Re}\{x_0\}=0$ and ${\rm Im}\{x_0\}$ steadily decreases to zero at $\rho_0=r_1$. In the interval $r_1<\rho_0\leq r_2$, we have constant ${\rm Im}\{x_0\}=0$ while ${\rm Re}\{x_0\}$  steadily increases to $K(m)$. Then for $r_2<\rho_0\leq r_3$, we have constant ${\rm Re}\{x_0\}=K(m)$ while ${\rm Im}\{x_0\}$ steadily decreases from zero to $-K(m')$. Beyond that, when $\rho_0\geq r_3$, the imaginary part remains constant ${\rm Im}\{x_0\}=-K(m')$ while the real part asymptotically approaches zero.

We can now examine how this plays into the density solutions with $\pm g$. For positive $g$, with all real roots, we know from our discussions earlier that physical solutions require $r_1\leq \rho_0\leq r_2$ (oscillatory, loop) or $\rho_0\geq r_3$ (non-oscillatory, wing).  In the oscillating region we find that ${\rm Im}\{x_0\}=0$  the $x_0$ is just the phase shift of the ${\rm sn}^2$ function. On the wing, a nonzero ${\rm Re}\{x_0\}$ still corresponds to a phase shift, but the ${\rm Im}\{x_0\}$  being an odd multiple of $K(m')$ provides an alternate picture of the how the solution can become unbounded in that region. We use the identity for ${\rm sn}$ with a complex argument shown in Eq.~(\ref{eq.complexsn}) in Appendix~\ref{SecAppB:Elliptic} and use the expressions in Eq.~(\ref{eq.jacobiK}) for the Jacobi elliptic functions evaluated at $K(m)$, the ${\rm sn}$ to objain
\bn {\rm sn}(u-iK(m'),m)=\frac{\sqrt{m'}{\rm sn}(u,m)}{1-{\rm dn}^2(u,m)}, \en
where we denote $u=\sqrt{g(r_3-r_1)}\ x +{\rm Re}\{x_0\}$ and we need necessarily have ${\rm Im}\{x_0\}=K(m')$ to make the imaginary part vanish in  Eq.~(\ref{eq.complexsn}). Clearly the expression above is unbounded as  ${\rm dn}^2(u,m)\rightarrow 1$. Notably, unphysical negative densities are naturally excluded since  the range of dn for $0\leq m\leq 1$ is given by $\sqrt{1-m}\leq {\rm dn}(u,m)\leq 1$).

For negative $g$, Fig.~\ref{Fig1_functionshape} indicates that for real roots, solutions can lie only between $r_2$ and $r_3$. Technically there is a wing  $\rho_0\leq r_1$, but the density cannot be negative. In Eq.~(\ref{eq.jacobidens}), for negative $g$, the co-ordinate dependent part of the argument becomes imaginary. In the oscillating region between $r_2$ and $r_3$, the ${\rm Im}\{x_0\}$ corresponds to a phase shift while the constant real part actually ensures that the density is oscillating. This is not so obvious in Eq.~(\ref{eq.jacobidens}), since naively the function appears to oscillate between $r_1$ and $r_2$. We once again use Eq.~(\ref{eq.complexsn}) and Eq.~(\ref{eq.jacobiK}) from the Appendix to transform the elliptic  function to get
\bn {\rm  sn}(K(m)+iv,m) = \frac{{\rm  dn}(v,m')}{1-m'{\rm sn}^2(v,m')} \en
Here $v=\sqrt{|g|(r_3-r_1)}\ x +{\rm Im}\{x_0\}$  and we need necessarily have ${\rm Re}\{x_0\}=K(m)$ for the imaginary part  to vanish in  Eq.~(\ref{eq.complexsn}).  This however still does not alter the limits of oscillation in Eq.~(\ref{eq.jacobidens}). We can remedy that with some additional identities and transformations for the Jacobi elliptic functions detailed in the Appendix~\ref{SecAppB:Elliptic}, we can actually transform the solution in  Eq.~(\ref{eq.jacobidens}) for negative $g$ to take a more transparent form
\bn \label{eq.dens2} \rho (x) = r_3+(r_2-r_3){\rm sn}^2\left(\sqrt{|g|(r_3-r_1)}\ x +\tilde{x}_0|\tilde{m}\right)\
 \en
where  $\tilde{x}_0=K(m)+{\rm Im}\{x_0\}$ is a real phase shift and $\tilde{m}=\frac{r_3-r_2}{r_3-r_1}$

This expression has intuitive consistency: In the case of positive $g$, we know an oscillating solution takes the form of Eq.~(\ref{eq.jacobidens}) and that $x_0$ is entirely real, corresponding simply to a phase shift in the density function. We can make the argument that $g \rightarrow -g$ simply reflects the function $f(\rho)$  across a vertical line through the middle root $r_2$ , so that $\{r_1,r_2,r_3\}\rightarrow \{r_3,r_2,r_1\} $ so that $r_1$ lies on the wing  and loop occurs between $r_2,r_3$. We see that swapping $r_1\leftrightarrow r_3$ in Eq.~(\ref{eq.jacobidens}) yields exactly the same expression as in Eq.~(\ref{eq.dens2}).

Similar arguments can be made for the complex roots corresponding to conjoined solutions. The satisfying and perhaps surprising conclusion here is that the general form of the solution Eq.~(\ref{eq.jacobidens}) works for all scenarios, regardless of the sign of $g$; and $x_0$ plays an essential role in determining whether the density is on a bounded or unbounded branch. Other solution forms used \cite{A14,A15} can be reduced to this by simply allowing for complex-valued $x_0$. A important point to be stressed in this context is that regardless of the sign of $g$, the imaginary and real parts of $x_0$, respectively, cannot be chosen arbitrarily. For positive $g$, $Im(x_0)=0$ corresponds to an oscillating solution, and $Im(x_0)=-K(m')$ corresponds to a wing solution. For negative $g$, $Re(x_0)=K(m)$ gives oscillating solutions, with no other feasible solutions.

\begin{figure}[t]
\centering
\includegraphics[width=\columnwidth]{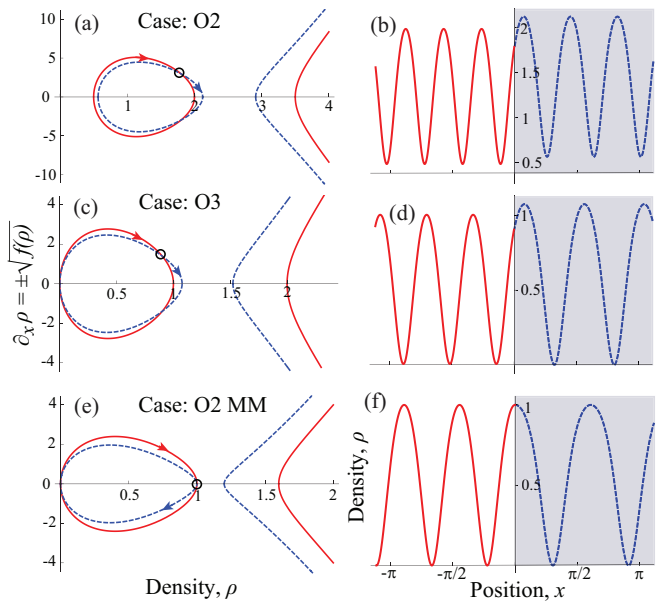}
\caption{(Color online) Distinct classes of solutions for a positive potential step $V_0>0$ with positive nonlinearity $g>0$, that are  \emph{oscillating} (O) in all regimes.  Left panels show phase space plots with red solid line for $f_L$ and blue dashed line for $f_P$, with the intersections corresponding to the potential boundary marked by circle.  Right panels show the corresponding density profiles, with the region with the potential shaded grey.  The most general scenario with two intersections O2 is shown in (a,b), and more restricted ones with three intersections O3 in (c,d) and with two intersections O2 MM at both the minimum and the maximum in (e,f). }
\label{Fig4_Vpos_gpos_bounded}
\end{figure}

\section{Step: Positive Nonlinearity}

The potential step $V_0$ for a medium with positive nonlinearity ($g>0$) presents the broadest set of possible solutions. Using our convention in Sec.~\ref{SecIV:Boundary}, the step boundary is at $x=0$, with $V_0>0$ in region $P:x\geq 0$, and $V_0=0$ in region $L:x<0$.  In a scattering problem, regions with no potential necessarily have bounded solutions, since they can extend to infinity. In those regimes, we will assume oscillatory solutions apart from some limiting cases we will consider separately.  Within the potential we will also consider solutions that are unbounded as $x\rightarrow \pm \infty$, because of their relevance for finite width potentials, even though unphysical for an infinite extent potential step. In the context of phase space plots, $\rho_L(x)$ will therefore always lie on a loop or asymptotic structure, while  $\rho_P(x)$ may lie on a loop, a wing, or a conjoined curve. The last two cases contain solutions unbounded at infinity. We will determine the distinct classes of allowed solutions by representing the boundary conditions as intersections between different types of phase space curves in the two regions.

\subsection{Oscillatory within the potential}
We first consider solutions that are oscillatory everywhere, so that the density lies on the loop portion of the phase space curves for both regions, and all the roots of $f(\rho)$ are ${r_i}\geq 0$. As discussed in Sec.~\ref{SecIV:Boundary}, the phase space curves of the two regimes can intersect at a maximum of three points. Due to the symmetry across the $\rho$-axis on the phase space plots, intersections will generally come in pairs, $(\rho_0,\pm\sqrt{f_0})$. The exception is when $\rho_0$ is a common root for $f_L$ and $f_P$ in which case $f_0=0$ and that would correspond to the minimum or maximum of the loop. Based on this, we have the following distinct types solutions, that we label with `O' for oscillating, the number of intersections, and indicate whether they occur at a minimum or maximum: O1 Min, O1 Max, O2, O3, O2 MM, where MM $\equiv$ Min Max.

Figure.~\ref{Fig4_Vpos_gpos_bounded} illustrates the last three cases. The most common case is O2 when there are only two intersections shown in panels (a,b), that correspond to oscillatory solutions of different amplitudes but none with nodes.  Cases O1 Min, O1 Max are not shown since they are limiting cases of O2, when $f_0=0$ and the two intersections shown in panel (a) merge into a single point of intersection that coincides with one of the extremes of the phase space loops. In panel (b), the boundary density would then correspond to a minimum or a maximum. The next most restricted is the case O3 shown in panels (c,d), where the loops intersect at three points with one of them necessarily being at zero density. But, as noted in Sec.~\ref{SecIV:Boundary}, in this case, the density at the boundary cannot vanish $\rho_0\neq 0$, so one of the intersections $\pm f_0\neq 0$ marks the boundary between the regions.

\begin{figure}[t]
\centering
\includegraphics[width=\columnwidth]{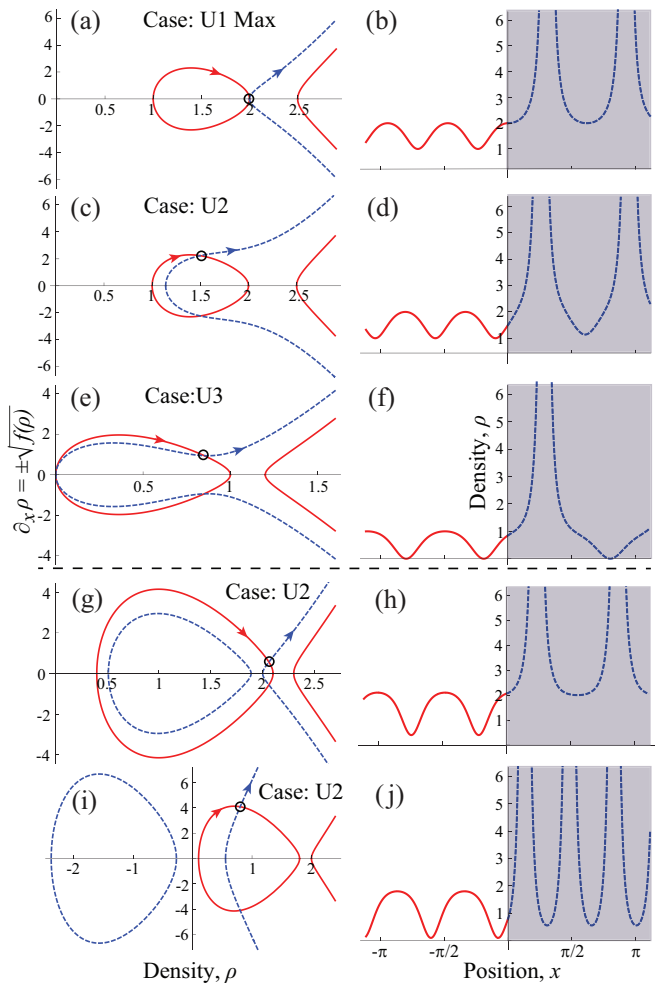}
\caption{(Color online) Similar to Fig.~\ref{Fig4_Vpos_gpos_bounded}, but showing the distinct cases with \emph{unbounded} (U) solutions in the region with the potential. (a-f) Scenarios where $f_P$ has a pair of complex roots and (g-j) scenarios where intersections occur on the wing segment of $f_P$. For complex roots, panels (a,b) correspond to U1 Max with intersection at the maximum point of the $f_L$ loop, (c,d) the general case U2 of two points of intersection and (e,f) the special case U3 of three points intersections. For intersections on the wing of $f_P$, the panels (g,h) show a general case U2 with  two intersections and all roots positive and (i,j) when $f_P$ has one positive and two negative roots.}
\label{Fig5_Vpos_gpos_unbounded}
\end{figure}

The most constrained oscillating solution O2 MM occurs when the loops intersect at zero as well as their mutual maximum, shown in panels (e,f). This can be viewed as a limiting case of O3, when the two intersections at $\pm f_0$ merge at the maximum of the two loops. Both O3 and O2 MM cases have nodes and therefore cannot carry current.  The latter is particularly special because the density oscillations have the same amplitudes in the $L$ and $P$ regions, but generally different shapes, as can be seen in panel (f). This makes it useful for comparing the effective potential ($V_0+g|\Psi|^2$). Note if the loop for $L$ has a zero root, current conservation ensures that there has to be an intersection there since the loop for $P$ has to have a zero root as well, as noted at the end of Sec.~\ref{SecIII:Constraints}.

\subsection{Unbounded within the potential}
We now examine another large class of solutions where the density is unbounded $\rho_P(x)\rightarrow \infty$ as $x\rightarrow \infty$ which were previously considered only for certain limiting cases \cite{A15} or omitted entirely \cite{A0,A7}. With oscillatory behavior in the $L$ region, there are two subclasses of unbounded density profiles in the $P$ region: Those that lie on a conjoined curve corresponding to a single real root, and those that lie on a wing while a loop exists as well. Distinct solutions of both types are shown in Fig.~\ref{Fig5_Vpos_gpos_unbounded}.

In a conjoined curve $\rho_P(x)$ has only one real root which has to be non-negative.
As in the previous subsection, the loop in the $L$ region can intersect the conjoined curve of the $P$ region at one, two or three points, and we label the solutions accordingly with `U' denoting unbounded: U1 Min, U1 Max, U2, U3.  With one of the curves being open, the case of two point intersections at both minimum and maximum of a loop is clearly not possible in this case. The case of one intersection at the maximum of the loop U1 Max is shown Fig.~\ref{Fig5_Vpos_gpos_unbounded}(a,b), and that with two intersections where $\rho_0$ lies at some intermediate value on the loop is shown in panels (c,d). The case of a single intersection at the minimum U1 Min is simply a limiting case of this and not shown. A case with three intersections U3, is possible as in the previous subsection, when the single real root of the conjoined curve is at zero, as shown in panels (e,f), and for the same reasons mentioned in the previous subsection, the density at the boundary $\rho_0\neq 0$ for this case. Qualitatively, the density profiles for all of these solutions are the same except for the U3 case, when solutions have a node.

The second class of unbounded solution have the density on the wing when a loop is present as well. Solutions with two intersections U2 are possible both when all the roots are positive as shown in  Fig.~\ref{Fig5_Vpos_gpos_unbounded}(g,h), and when there are two negative roots and one positive root shown in panels (i,j). U1 Max solutions can occur as a limiting case of the former and U1 Min solutions as a limiting case of the latter, when $f_0=0$ corresponds to the sole non-negative root of $f_P(\rho)$. However U3 solutions where all three intersections are on the $L$ loop analogous to panel (e) do not seem to be available, since as the $P$ wing gets closer to the zero it tends to get steeper and less likely to intersect again with the $L$ loop. In principle, the loops from $L$ and $P$ could meet at zero in a variation of that plot with the wing still intersecting at two points, yielding a U3 case, but with the $P$ loop in the unphysical negative density regime, it is not relevant.

\subsection{Degenerate Roots}
\begin{figure}[t]
\centering
\includegraphics[width=\columnwidth]{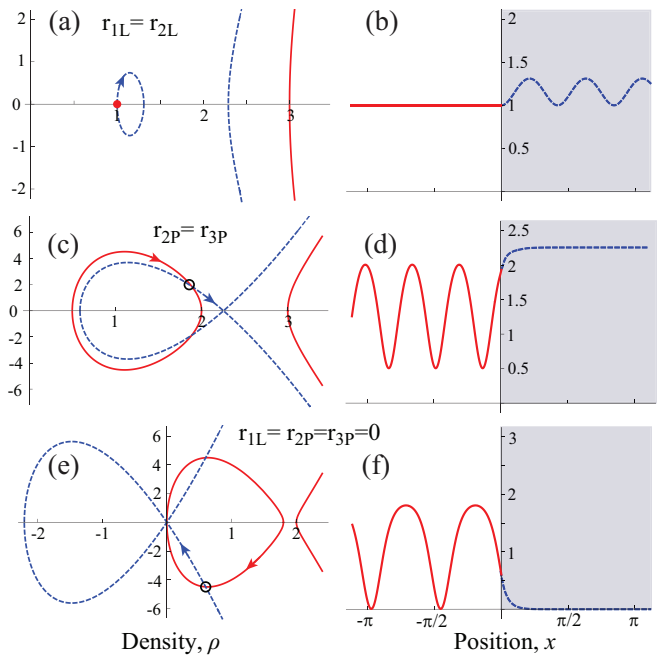}
\caption{(Color online) Similar to Fig.~\ref{Fig4_Vpos_gpos_bounded}, limiting cases are shown where two roots are degenerate for either $f_L$ or $f_P$. Panels (a,b) show the case of $r_{1L}=r_{2L}$ for $f_L$, yielding a plane wave, reverse is possible as well with plane wave in the potential with $r_{1P}=r_{2P}$ for $f_P$. (c,d) $r_{2P}=r_{3P}\neq 0$ and the intersection occurs on the loop part of $f_P$ the density approaches finite constant value. (e,f) $r_{2P}=r_{3P}=0$ for $f_P$ with the relevant intersection on negative derivative $\rho'<0$ of the wing leading to a decay solution.}
\label{Fig6_Vpos_gpos_limiting}
\end{figure}

When any two roots are equal, we obtain limiting cases of oscillating or unbounded solutions discussed above, depending on where intersections occur on the phase space curve of $P$. Relevant cases are shown in Fig.~\ref{Fig6_Vpos_gpos_limiting}.

When the lower two real roots are equal $r_1=r_2$, the solutions are plane waves, as shown in panels (a,b). The plane wave can be in either region, here we show it to be on the $L$ side. Clearly the derivative at the edge of the potential must be zero, hence these solutions can be considered the limiting cases of O1 Min and O1 Max cases, where $L/P$ collapses to a point that coincides with the minimum or the maximum of the $P/L$ loop.

When the two upper  roots are degenerate $r_2=r_3$, we have asymptotic solutions that approach a constant value away from the step boundary.  Since such solutions are by definition bounded, they can occur in all segments of a phase space curve including wing or conjoined, and they are valid for an infinite extent step potential. Asymptotic solutions can occur in either region, with or without a potential. An example is shown in panels (c,d) with the degeneracy within the $P$ region.  Figure~\ref{Fig2_fp_minus_fl} implies that increasing the potential brings the loop and the wing closer together to create the degeneracy. This determines what type of density solutions across the step would be compatible with such asymptotic behavior. For $g>0$ and $V_0>0$, if $\rho_L$ is an asymptotic solution on either the loop or the wing, $\rho_P$ will be unbounded. If $\rho_P$ is an asymptotic solution on either the loop or the wing, $\rho_L$ must be an oscillating solution.

When the $L$ loop intersects the $P$ curve on the loop section, as shown in panel (c), the clockwise circulation along the curve ensures that the density approaches the asymptote from below. That also is the reason why the density approaches a constant asymptotic value: Once it gets to the $X$-shaped part in the $P$ curve, it cannot continue on the upper right or lower left branches because they are discontinuous transitions within the same region, and the only smooth transition to the lower right branch is forbidden by the clockwise consideration because a negative $\rho'$ cannot lead to increasing density. On the other hand, if in panel~(c) the $L$  curve extends farther right and intersects on the wing of the $P$ curve, then for exactly the same reasons the density would approach an asymptotic value from higher values, following the lower right branch of the $X$-shaped part of the $P$ curve.

Equation (\ref{parameters}) shows that for asymptotic solutions, the elliptic parameter $m=1$, so that wavelength goes to infinity implying no periodicity. Asymptotic solutions would therefore correspond to gray or dark solitons \cite{B20,B4,B5,A10,A11,A12,A13,Hakim-NLSE_delta_potential,B21_1,B21_2,A15}, where in a uniform media one would just have the $P$ curve with a pair of degenerate roots in panel (c). The asymptote in the $P$ region can be extended to the left creating a dip before it forms another asymptote on the left, creating the dark soliton. In case of a step potential, depending on whether the $L$ curve intersects the $P$ curve on the loop or the wing side, these solutions can be considered limiting cases of O2 or U2 solutions. Only the O2 cases would be allowed in an uniform medium, with U2 leading to diverging solutions.

A special case of the above can occur when in panel (c) the $L$ curve intersects the $P$ curve exactly at the degeneracy point, in case the asymptotic solution within the potential would become a plane wave. Of course $L$ an $P$ regimes can be switched to have the plane in the no potential regime. Such solutions are relevant for example in Ref.~\cite{A15}, where the degenerate solutions occur outside the potential and unbounded within it.

When the degeneracy occurs at zero density, $r_2=r_3=0$ and the $P$ loop lies in the negative density regime, we have decaying solutions that approach zero asymptotically, as shown in panels (e,f). For $g>0$, the decay can only occur on the side of the higher potential, because as mentioned above $r_2,r_3$ approach each other to create the required merger as the potential increases. This is simply an asymptotic solution with a vanishing limit. Comparing to Fig.~\ref{Fig5_Vpos_gpos_unbounded}(i,j) this can also be viewed as its limiting case with three intersections instead of two, $U2\rightarrow U3$.

Decay, asymptotic, and plane wave solutions are all bounded solutions with similar features. They differ in that: Intersections for decay solutions  will never occur at a minimum and require a zero root, intersections marking $\rho_0$ for asymptotic solutions never occur at whichever extremum the density asymptote, and plane wave solutions will only have intersections where the derivative vanishes.

\begin{figure}[t]
\centering
\includegraphics[width=\columnwidth]{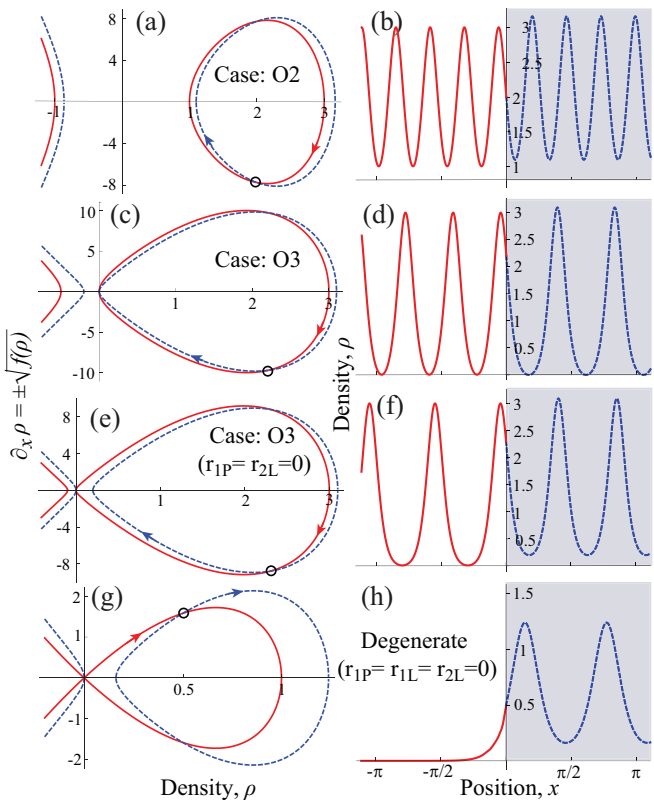}
\caption{(Color online) Similar to Fig.~\ref{Fig4_Vpos_gpos_bounded}, solutions for a positive potential step $V_0>0$ with \emph{negative} nonlinearity $g<0$, are shown for (a,b) a general two-intersection case O2, (c,d) the special case O3 with three intersection in (c,d). Figures (e,f) shows a case where $\rho_L(x)$ has nodes but $\rho_P(x)$ does not, impossible with $g>0$. (g,h) A solution that decays on outside the potential, in contrast with Fig.~\ref{Fig6_Vpos_gpos_limiting}(c,d) for $g>0$ where that can only occur within the potential as with the linear case.}
\label{Fig7_Vpos_gneg}
\end{figure}

\section{Step: Negative Nonlinearity}\label{SecIX:Step:Negative}
With negative nonlinearity, $g<0$, as discussed at the end of Sec.~\ref{SecIII:Constraints}, complex roots are not allowed and the wing part of the phase space curve opens on the left and lies in the unphysical negative density regimes. So, for a potential step, the only allowed solutions possible, in any region $L$ or $P$, have to lie on the loop part of the respective phase space curves. Solutions of type O2 with two intersections are shown in Fig.~\ref{Fig7_Vpos_gneg}(a,b), and clearly type O1 Min, O2 Max and O1 MM solutions are allowed as limiting cases similarly to the $g>0$ case.  Solutions with three intersections, O3, are shown in panels (c,d), where the middle root $r_2=0$.

The most interesting case for $g<0$, and one not possible with $g>0$, is shown in panels (e,f) when solutions can have nodes in the $L$ regime but are nodeless in the $P$ regime. Here, we do not have to obey the rule as with $g>0$ that the densities in all regions must have a node if one region has a node. We can still satisfy current conservation condition, mentioned at the end of Sec.~\ref{SecIII:Constraints}, which mandates that if one region has a zero root all regions need have at least one, because here within the potential the zero root is on the wing $r_{1P}=0$, while the left side has zero root at $r_{2L}=0$.

Although not shown here, comparison with Fig. ~\ref{Fig6_Vpos_gpos_limiting}(a,b) confirms that plane waves in either regime are also possible, since it simply requires the two highest roots to be degenerate $r_2=r_3$, such that one or both of the loops shrink to a point. However asymptotic solutions that require the loop and the wing meet at a degeneracy point $r_1=r_2$  can only occur when density decays $\rho\rightarrow 0$ because the wing and the loop lie in negative and positive density regimes, and they can only meet at $r_1=r_2=0$.  Such decay solutions may exist in any potential region, provided it transitions to bounded solutions across the potential boundary; an example with decay occurring in the $L$ region is shown in Fig. ~\ref{Fig6_Vpos_gpos_limiting}(g,h). Here $r_{1L}=r_{2L}=0$, but to satisfy current conservation $r_{1P}=0$.

With negative nonlinearity, these decay solutions would correspond to a single bright soliton \cite{A10, A11, A12, A13, B4, B20, B21_1}: For example, for a medium without a potential in panel (g), if we follow the entire loop in the $L$ region, we would have the density profile of a single bright soliton, with the return to $(\rho, \rho')=(0,0)$ marking the asymptotic approach to vanishing density in the other  direction.

\section{Rectangular Potential Barrier}

\begin{figure}[t]
\centering
\includegraphics[width=\columnwidth]{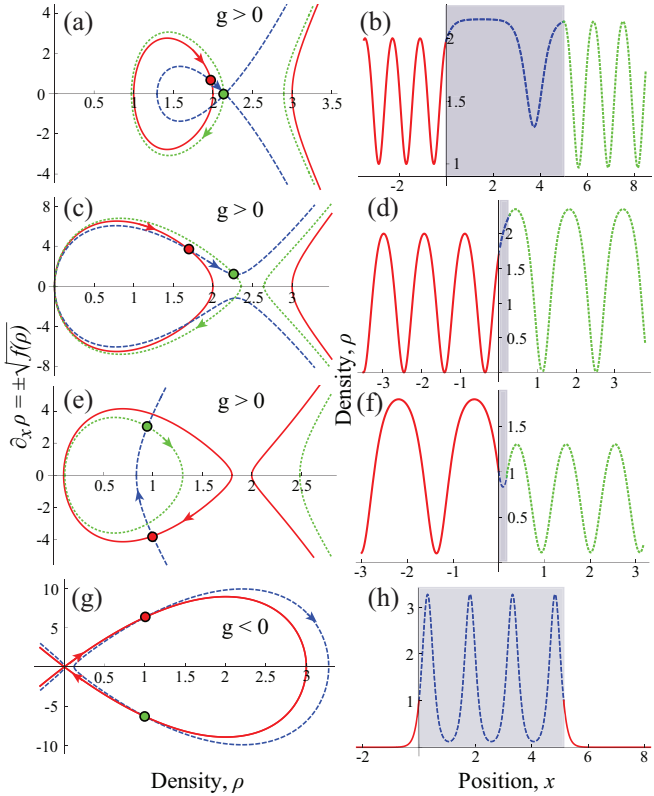}
\caption{(Color online)  Examples of distinct solutions for a rectangular barrier are shown, with left panels showing the phase space curves on the left of (solid red), within (dashed blue) and on the right of (dotted green) of the barrier shown as gray shaded in the right panels that plot the density profiles. Panels (a-f) have $g>0$ and assume oscillating density, lying on the loop segment of the phase space curve. Within the barrier, density can be (a,b) on the loop, (c,d) the conjoined form or (e,f) on the wing. The two latter cases can blow up and set strong constraints on the width. (g,h) For $g<0$ a bound state can exist within the barrier, decaying outside; L/R curves coincide in (g) due to optional symmetry assumed.}
\label{Fig9_barrier}
\end{figure}

We can construct solutions for the rectangular barrier using the solutions we have found for the $L-P$ step and their mirrored counterparts for the $P-R$ step at the right edge of the barrier. The complete landscape of solutions discussed above for a step potential are physically relevant for a finite width barrier. In Fig.~\ref{Fig9_barrier}, we present some examples. With three regions, there are three phase space curves, depicted with solid red on the left ($L$) of the barrier, dashed blue within the potential ($P$) and dotted green for the right ($R$) of the potential.  There are now two points of the intersections marking the $L-P$ and the $P-R$ boundaries respectively.  The density at the right edge of the barrier at $x=a$ is denoted $\rho_a$ in analogy with $\rho_0$ on the left edge, and $a$ is the width of the barrier.

Panels (a-f) assume $g>0$  and oscillatory solutions outside the barrier corresponding to loop segments of the phase space curves, but illustrates cases for the density lying on different phase space segments within the barrier: On a loop in (a,b), on a conjoined curve in (c,d) and a wing in (e,f) when a loop is present (in this case, on the negative density regime as in Fig.~\ref{Fig5_Vpos_gpos_unbounded}(i)). Outside the barrier, we can also have plane wave or asymptotic solutions, however we cannot have decay solutions since they only occur in the region of higher potential for $g>0$, as discussed in the context of Fig.~\ref{Fig6_Vpos_gpos_limiting}(i,h).

For negative nonlinearity $g<0$, since the densities have to lie on the loop segment of the phase space curves in all regions, all the various types of oscillatory solutions described in Sec.~\ref{SecIX:Step:Negative} are possible here. The general behavior and appearance can be surmised from our discussion in the section. Therefore, in Fig.~\ref{Fig9_barrier}(g,h) we only show the one exception to this, and hence the most interesting case, where the solution decays outside the barrier. Such localization within a region of higher potential is a curious effect of the negative nonlinearity and cannot occur in a linear system or for positive nonlinearity.

There are however certain constraints that arise with a finite width barrier. The primary one is that for those solutions that are unbounded at infinity, the barrier needs to be sufficiently narrow to intercept only a finite-valued segment of the density. As apparent from Fig.~\ref{Fig5_Vpos_gpos_unbounded}, for such unbounded solutions the density blows up periodically, and the width of the barrier has to be less than half of the period. Secondly, since there is only one point with a zero derivative on wing or conjoined features of the phase space curves, $L$ and $R$ solutions cannot \emph{both} have extrema at the boundary if they are connected by an unbounded solution within the potential.

\begin{table}
\centering
\includegraphics[width=\columnwidth]{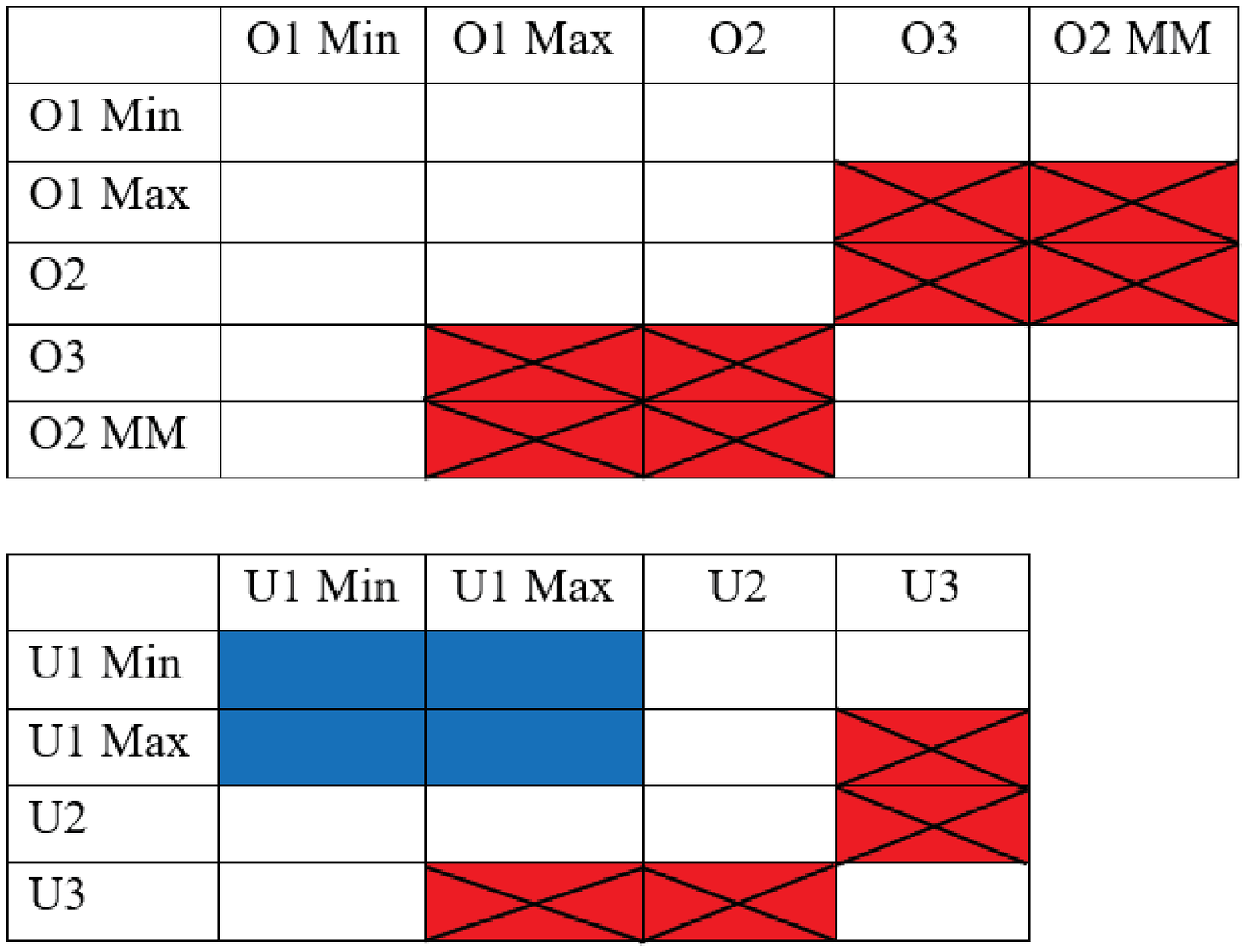}
  \caption{This table shows the allowed combinations of solutions at the left (vertical labels) and right step edges (horizontal labels) of a rectangular barrier. The labels O and U stand for oscillatory and unbounded solutions within the potential. The number represents intersections between the phase space curves at each edge, and tags specify if they occur at a minimum, maximum, or both (MM). The blue shaded cells mark solutions that are not possible due to the density reaching infinity before reaching the other edge, and the red crossed cells mark solutions not allowed by current conservation. Decay, asymptotic, and plane wave solutions are limiting cases of those listed, when two of the roots are degenerate.}
  \label{Fig10_Table}
\end{table}

Table~\ref{Fig10_Table} depicts all allowed combinations of solutions, using the labelling we introduced.  The solutions outside the barrier must all be bounded, hence the labels $O$ or $U$ here indicate the solution type within the potential region $P$. The numerical labels and the other tags indicate number of intersections of the phase space curve of $P$ region with those of the adjacent $L$ and $R$ regions. The column labels represents the relevant $P$ solution type and phase space intersection on the left edge of the barrier, and the row labels, the right edge of the barrier.

The table shows that even when all these solutions for a step potential satisfy the boundary conditions at on edge, there are restrictions on those solutions they can pair up with on the other edge. The blue shaded regions indicate combinations that are forbidden because the density will go to infinity before having meeting the required boundary condition at the other edge. The red shaded boxes with crossed out cells in the table mark combinations that are forbidden by current conservation, where there would be a zero root on one side of the barrier but no zero root on the other side. The rectangular barrier is symmetric, so the solutions put together in this grid are symmetric across the main diagonal. For example, the element connecting O1 Min to O1 Max is the same solution as the element connecting O1 Max to O1 Min, just mirrored about the center of the barrier. The table does not explicitly list solutions with degenerate roots, since those are limiting cases of the solutions shown.

\begin{figure}[t]
\centering
\includegraphics[width=\columnwidth]{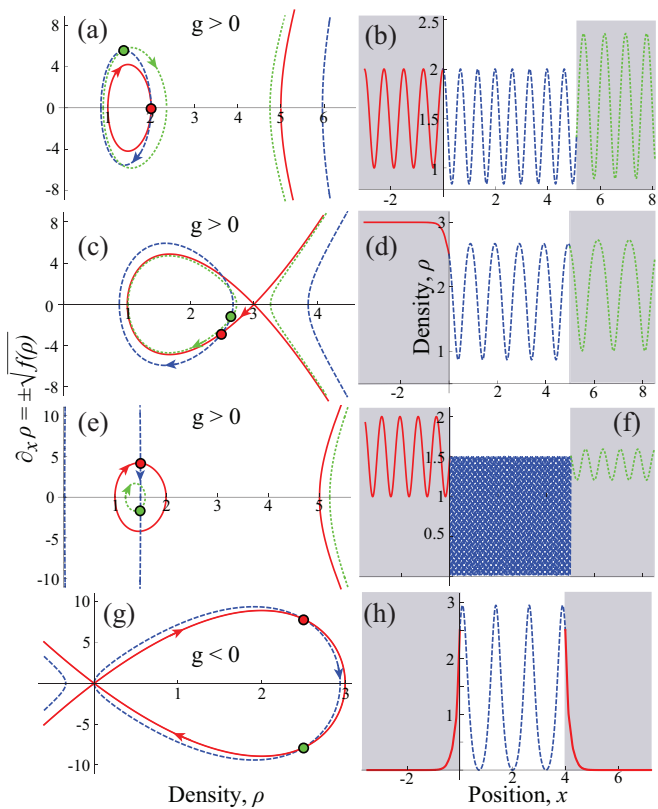}
\caption{(Color online) Similar to Fig.~\ref{Fig9_barrier}, but for potential well with $V_0<0$ in $0<x<a$, shown as \emph{un}shaded region in right panels. The key difference with a barrier is that only bound solutions are permitted regardless of the sign the of nonlinearity. For $g>0$, example of a solution oscillatory everywhere is shown in (a,b) and asymptotic outside the well in (c,d). In the limit of a deep well $|V_0|\rightarrow \infty$ and $g>0$ the density oscillates with increasing frequency between 0 and $\rho_0$ (e,f). Bound states, decaying outside the well exists for both positive and negative nonlinearity, shown in (g,h) for $g<0$ with optional L/R symmetry.}
\label{Fig11_well}
\end{figure}

\section{Rectangular Potential Well}
When we flip the sign of the potential, changing from a barrier to a well, $V_0<0$ in the $P$ region, we switch to stepping up in the potential value on the right edge stepping down on the left edge. This alters the types of solutions allowed compared to the potential barrier.

The case of attractive nonlinearity, $g<0$, is straightforward, since one root must lie in the unphysical negative density regime which also corresponds to the wing part of the phase space curve. Therefore, the allowed density solutions are necessarily bound solutions that lie on the loop segment of the phase space curves, and allows only oscillatory or decay solutions. The behavior is basically the same as for a potential barrier. So, the main takeaway for $g<0$ is that sign of the potential does not alter the landscape of solutions available.

With repulsive nonlinearity, there are two main considerations: First, comparing Fig.~\ref{Fig2_fp_minus_fl}(a) and (c), changing the sign of $V_0$ leads to opposite migration of the roots, because the quadratic function $\Delta f$ in Eq.~(\ref{eq_f}) has opposite concavity. Second, and more crucially, Fig.~\ref{Fig3_roots_migration}(c) shows that for $g>0$, as $V_0$ increases the solutions transition from being oscillatory (region I in that figure) to unbounded (regions II and III), with the boundary between regions I and II marking decay and asymptotic solutions. This means that unbounded solutions will not be allowed within a potential well since $V_0$ increases outside the well and that would mean physically impossible unbounded solutions in the $L$ and $R$ regions as well. In fact, the solutions within the well can only be oscillatory in nature, since if the solution lies on the boundary of regions I and II as mentioned above, the increase in the potential outside the well will tip the solutions over to the unbounded regime II. This is clearly not an issue for a potential barriers, since the potential decreases outside the barrier, and solutions that lie in regions II and III within the barrier can transition to bounded solutions in or on the edge of region I.

An example of a generic solution for a potential well, which is oscillating in all regimes is shown in Fig. ~\ref{Fig11_well}(a,b).  However, the conditions above do still allow plane wave, decay and asymptotic types \emph{outside} the well, an example with an asymptotic solution on the left of the well is shown in Fig.~\ref{Fig11_well}(c,d).

Differently from a potential barrier, a well can support bound states for both positive and negative nonlinearities. They arise the same way as with the potential barrier. For $g<0$, outside the well, we need to have degenerate roots $r_1=r_2=0$ and a positive third root $r_3>0$.  An example is shown in Figure ~\ref{Fig11_well}(g,h), where decaying densities are symmetric in the $L$ and $R$ regimes but such symmetry is not necessary. For $g>0$, similar bound states exist within the well, oscillating within and decaying outside the well, provided there are degenerate roots outside the well $r_2=r_3=0$ and some negative first root, $r_1<0$. These class of bound state solutions include localized solutions studied in Ref.~\cite{A5}.

Something interesting occurs uniquely for a potential well with repulsive nonlinearity $g>0$: When the well depth $V_0\rightarrow -\infty$, density at both boundaries match $\rho_0\simeq \rho_a$ as seen in  Fig.~\ref{Fig11_well}(e,f), and the mean value of the density approaches a finite limit $\rho_0/2$ with the density oscillating between $0\leq\rho\leq\rho_0$ with increasingly higher frequency. This is illustrated in Fig.~\ref{Fig11_well}(e,f) for sufficiently large $|V_0|$. This cannot occur for $g<0$, since as the well depth is increased, the roots migrate to values that are not physical as discussed in Sec.~\ref{SecIII:Constraints}. For potential barriers, $V_0>0$ with repulsive nonlinearity as the potential increases the density becomes unbounded as we concluded from Fig.~\ref{Fig3_roots_migration}(c), while with attractive nonlinearity the amplitude of density oscillations blows up.

\begin{figure}[t]
\centering
\includegraphics[width=\columnwidth]{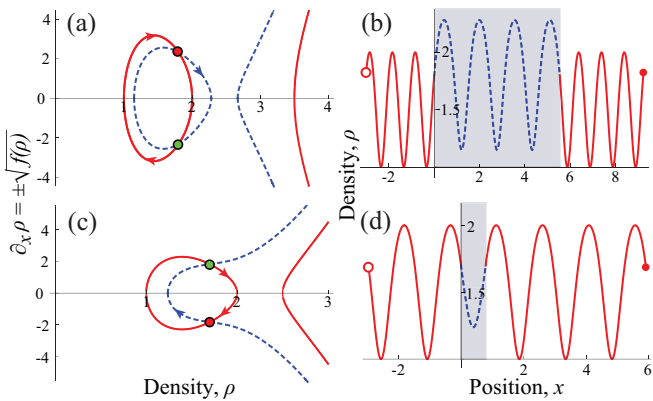}
\caption{(Color online) Examples of solutions for a barrier on a ring topology with (a,b) a general case of oscillating density within the potential  (c,d) an unbounded solution within the potential shown here for conjoined case, but applies also to a wing. There are only two regions, inside and outside of the barrier, and hence only two phase space curves in (a,c) and the same density at both edges (b,d) which requires the unbounded densities (c) in the barrier have a negative derivative on the left edge.}
\label{Fig12_ring_solutions}
\end{figure}

\section{Barrier Potential on a Ring}\label{Sec:Ring}
Changing the boundary conditions to have a barrier potential in a finite ring topology introduces some interesting changes in the solutions. Clearly, only oscillating solutions are relevant for outside the barrier, since unbounded ones are not possible and decay solutions would not be significantly impacted. Instead of three regions, the potential, its left and its right, on a ring there are only two regions: a single region outside of the potential and the region inside. Thus, for a symmetric barrier, the density at both edges must be equal to $\rho_0$, and the derivatives must match, up to a sign. This can be understood in terms of the phase-space plot in Fig.~\ref{Fig12_ring_solutions}, there being only one loop (solid red line) for outside the potential, its intersections with the loop or wing for within the barrier (dashed blue line) is the same for both edges apart from the choice of being above or below the $\rho=0$ axis. Examples of analytical solutions for a ring topology have been previously examined, but only in terms of oscillating solutions in each region \cite{A1,A2}.

For any given barrier width, there is at least one solution which fits this criteria. For oscillating solutions, this means that there is a solution with either an integer number of wavelengths which matches the width of the barrier, or some number of wavelengths which intersects at the edge of the barrier at the same value $\rho_0$ but at the opposite signs of $\rho'$ at the two edges. The latter case is shown in Fig.~\ref{Fig12_ring_solutions}(b). In the case of extremely narrow barriers, unbounded solutions are possible but must have $\rho'<0$ on the left edge and $\rho'>0$ on the right edge of the barrier and remain finite in between, as shown in Fig.~\ref{Fig12_ring_solutions}(d) . This is the same as our condition in the open case where oscillating densities hold for any barrier width, and imaginary and third root solutions hold only for a sufficiently small barrier width. Asymptotic solutions are not possible, as it does not satisfy the criteria that the density and its derivative must be the same on either side of the barrier.

There is a significant additional restriction for solutions on a ring. For open boundary conditions as we have considered so far, the phase has not been an issue, since for any well-defined the phase is simply defined by Eq.~({eq1.2-1}) with $\Omega=0$ and it is not a constraint. But, in a ring, the phase and its derivative must satisfy the periodic boundary condition in Eq.~(\ref{eq1.2-1}) \cite{Das-Huang}. Without rotation, a solution can only exist if the phase change around the ring satisfies $\Delta \phi=2\pi n$ with integer $n$. However, rotation provides a continuous parameter $\Omega$ that can be adjusted to meet the phase constraint for any solution that meets the density criteria specified above. But it is very relevant that the vast majority of those solutions will not be valid in the absence of rotation.

\section{Smooth Barrier}\label{Sec:smooth}

The choice of rectangular barrier or a step potential which are piecewise constant is dictated by the fact that we can find analytical solutions, for the same reason that they are considered in the linear case.  As with the linear scattering problem, such potentials capture the essence of scattering by more general potentials.  We now show that this is the case for the nonlinear problem as well.

For convenience of numerical simulation, we illustrate this with a ring potential, but the primary conclusions are generally applicable. The periodic boundary condition of the ring makes it convenient to use a momentum space analysis with a finite basis. We expand the state and the relevant potential as
\begin{equation}\label{momentum_space}
  \psi=\sum\limits_{m=-N_h}^{N_h} c_m e^{imx}\h{1cm}  V(\theta)=\sum_{n=-N_h}^{N_h}v_n e^{inx}
\end{equation}
so the time-independent nonlinear Schr\"odinger equation reduces to a set of $N=2N_h+1$ coupled equations
\begin{eqnarray}
    {\textstyle\frac{1}{2}}(m-\Omega)^2c_m+\sum_{n_1,n_2}v_{n_1}c_{n_2}\delta_{n_1+n_2-m}\n\\
    +g\sum_{n_1,n_2,n_3}c_{n_1}^* c_{n_2} c_{n_3} \delta_{n_2+n_3-n_1-m}=\mu c_m
\end{eqnarray}
We use a basis size of $N=51$ and solve the equations with a generalized Newton's method \cite{Das-Huang} to find the coefficients $c_n$ for a specific potential $V(x)$ and nonlinear strength $g$. As mentioned in the previous section, we allow for the appropriate rotation $\Omega$ essential to match the phase boundary condition, so that $\Delta\phi(x)=2\pi n$ for a complete circuit of the ring.

\begin{figure}[t]
\centering
\includegraphics[width=\columnwidth]{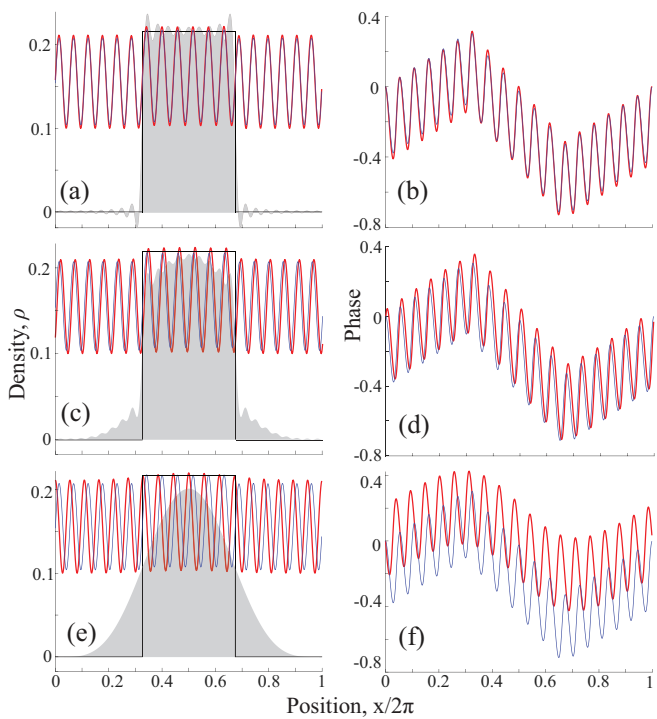}
\caption{(Color online) Numerical solutions (thick red lines) for barrier potential (shaded) on ring, morphing from a rectangular form constructed with its Fourier components (a,b) to a $\sin^4(x/2)$ profile of equal area in (e,f) with intermediate form 7:3 mixture of the two limits in (c,d). Left panels show the density and the right panels show the phase. The solutions are superimposed with the analytical solutions (thin blue line) for a rectangular barrier potential of the same dimensions, with $V_0=0.6$. The general features of the analytical solutions remain applicable to a smooth barrier as well. }
\label{Fig13_smooth_barrier}
\end{figure}

We find the solutions analytically for a rectangular barrier on a ring as in Sec.~\ref{Sec:Ring}, taking $V_0=0.6$, $g=5$, with the roots outside the potential $r_L=\{1;2;5\}$ and density at the boundary $\rho_0=1.5$ and its derivative $\rho'>0$. The ring size is specified by requiring the number of periods inside and outside the barrier to be fixed at $6$ and $12$ respectively.  For comparison with our numerical solutions on a ring we rescale the system length $L\rightarrow 2\pi$ and rescale all the parameters and the solutions in proportion and specifically ensure that the density is normalized to unity on the ring.

In the physics of ultracold atoms, localized barriers/wells can be created with tightly focussed blue/red detuned lasers which present a Gaussian profile \cite{ramanathan,Das-Aubin-PRL2009}. We use a similarly shaped potential of the form $V_{sin}=\sin^4(x/2)$, the sinusoidal form being convenient for our chosen basis. In order to determine whether the analytically obtained solutions for the rectangular barrier are also applicable for smooth barriers relevant in experiments, we gradually transform the rectangular potential $V_{rect}$ of dimensions used in the analytical simulation into the $V_{sin}$ potential by ramping up the weight from $0$ to $1$ in the composite potential
\begin{equation}
	V(x)=(1-w)\times V_{rect}(x)+ w\times V_{sin}(x),
\end{equation}
and determine the solutions numerically.  For relevant comparison, $V_{sin}$ is chosen so that the area of the sinusoidal potential equals that of the rectangular barrier. The results are shown in Fig.~\ref{Fig13_smooth_barrier}. The left column shows the density profile while the right column shows the phase of the solution. From top to bottom, the weights are $w=0, 0.3, 1$ respectively. The original analytical solution with rectangular barrier is always shown as thin blue lines for comparison. The numerical solution is overlaid in thicker red lines. The mixed potential $V(\theta)$ together with the rectangular barrier are also plotted in shaded profile. The finite sized basis introduces some wiggles in representing the rectangular potential but, as panels (a,b) confirm, the numerical solution matches the analytical solution well.

What stands out is that both the density and phase are not substantially altered in transitioning from the discontinuous rectangular barrier in Fig.~\ref{Fig13_smooth_barrier}(a,b) to the smooth sinusoidal barrier in Fig.~\ref{Fig13_smooth_barrier}. With regards to the density, the period remains the same but the profile is shifted; and the sudden upsurge of the mean density within the barrier in the rectangular case is transformed to a gradual ramping up and down across the smooth barrier. The phase also retains the same profile and follow the lateral shift in the density modulation. The more conspicuous vertical shift is effectively a constant offset that is physically irrelevant.

This comparison shows that the analytical solutions determined in this paper in the context of a rectangular barrier can be applicable to the broad range of localized potentials cosine potentials \cite{B10, B11} or Gaussian potentials typical of lasers in atomic physics.  Of course, any nonsingular potential can be approximated by a series of adjacent rectangular potentials, in the nature of a finite Riemann sum. Furthermore, even solutions for periodic potentials such as Kronig-Penney \cite{A4}, or optical grating \cite{B15} can build on these solutions for unit cells in conjunction with Bloch's theorem.

\section{Stability of Solutions}

We will now explore the dynamical stability properties of the solution by considering small perturbation around the mean field stationary states:
\bn \psi(x,t)=\psi_0(x)+\delta u e^{-i\mu t} e^{-i\omega t}+\delta v^* e^{-i\mu t} e^{i\omega^* t}.\en
We solve the resulting Bogoliubov equations \cite{RMP-Sringari-1999} for the normal modes of the fluctuations.
\bn (H_0+2g|\psi_0|^2-\mu)\delta u+ g\psi_0^2\delta v=w \delta u\n\\
-(H_0^*+2g|\psi_0|^2-\mu)\delta v- g(\psi^*_0)^2\delta u=w \delta u\en
where $H_0=\frac{1}{2}\left(-i\pt\right)^2+V$. The numerical solutions for the fluctuations $\delta u$ and $\delta v$ are done with same momentum state expansion used in  Eq.~(\ref{momentum_space}).  We find the normal modes for fluctuations by diagonalizing the resultant $2N\times 2N$ square matrix that arises from the Bogoliubov equations.  If the angular frequencies $\omega$ of the normal modes have positive imaginary components ${\rm Im}(\omega)>0$, then the fluctuations would grow exponentially, indicating dynamical instability. For purely imaginary $\omega$, the eigenstates of fluctuations $\delta u$ and $\delta v$ are identical, but typically not so for complex or real values.

\begin{figure}[t]
\centering
\includegraphics[width=\columnwidth]{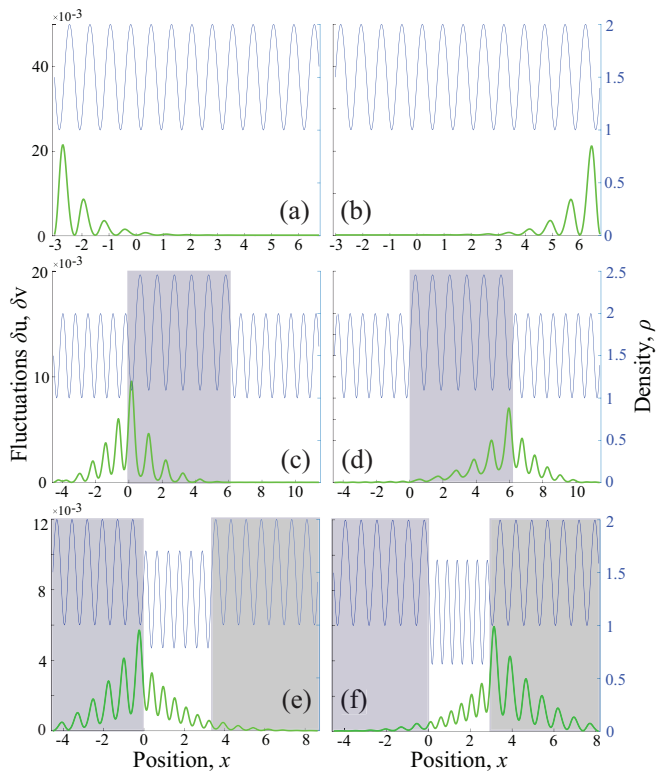}
\caption{(Color online) The eigenstates (thick green lines, left axis label) superimposed on images of the density (thin blue lines, right axis label). Instability is localized in places where the density derivative $\rho'$ at an edge indicates a trend opposite to the change of the actual mean value of the density. Examples are shown for no potential with instability localized at the edge of the system (a,b), a potential barrier at the barrier edges (c,d), and a potential well at the edges of the well (e,f).}
\label{Fig14_stability_states}
\end{figure}

We find that persistent instabilities appear primarily as a consequence of discontinuous edges and boundaries. Considering the infinite limit typically assumed in scattering problems, the boundary effects should be less relevant, but instabilities at potential edges still remain. Our main observation about the instabilities whether at a boundary or at a potential is as follows:
\emph{Solutions are unstable when the derivative of the density, at an edge of a potential or the whole system, has a sign opposite to the actual change in the mean value of the density across that edge}.  This is illustrated in Fig.~\ref{Fig14_stability_states} where we plot the unstable modes of the Bogoliubov equations.  For example, if the density has a positive derivative at an edge of the whole system, that solution will be unstable at that edge, since the mean value of the density abruptly decreases to zero outside the system. This can be seen in panels (a,b) of that figure, with instability localized at one edge but not at the other in each case. The remaining panels show similar behavior at the edges of the potential, with the instability localizing at that edge where the derivative of the density and change in the mean density are in opposition. For example in panel (c), on the left edge of the barrier $\rho'<0$ going into the barrier but the mean value increases within the barrier, so there is an instability localized there, while at the right edge the $\rho'<0$ going out of the barrier but the density decreases as well, so there is no instability localized there.

We now examine the trends in the instability as the height or depth of the barrier changes. In scattering problems, typically infinite systems are assumed.  But, for our simulations, we have to assume a finite system size, introducing system edges with accompanying instabilities as mentioned above.  Therefore in order to differentiate the effects of the boundary on the stability, we do a comparative study by progressing from a (i) closed ring with no boundary, to (ii) a cut ring, an open system where the system is adjusted such that the state at the two extremes match continuously both in density and phase, and finally to (iii) a general open system where the density and phase of the state at the two edges can be arbitrary and independent of each other.

\begin{figure*}[t]
\centering
\includegraphics[width=\textwidth]{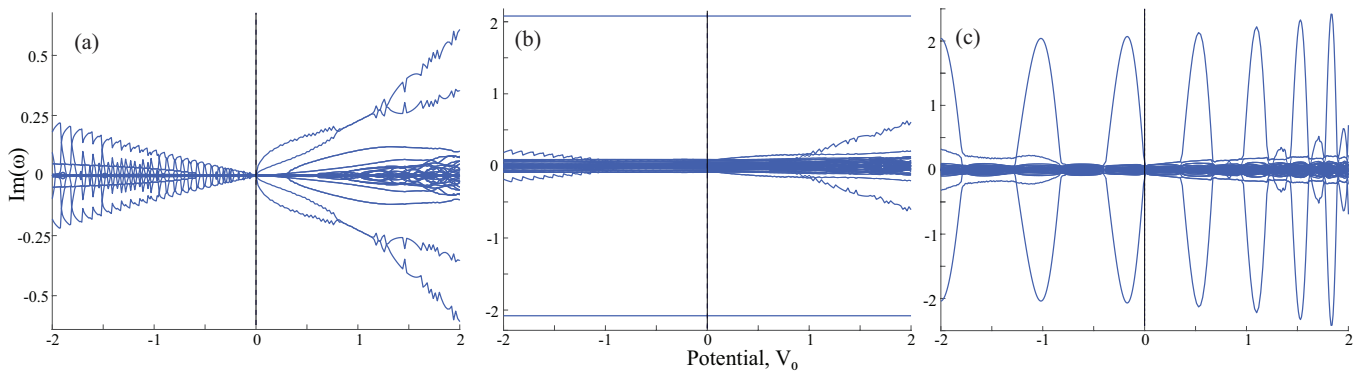}
\caption{(Color online) Instability of nonlinear scattering solutions of potential barrier/well as a function of the strength ($\pm V_0$) as gauged by the imaginary part of normal modes of Bogoliubov fluctuations. The behavior is shown for (a) ring boundary (b), an intermediate cut ring case with box boundary condition but matching density and phase at system edges and (c) box boundary condition. For all, the bulk of the modes are complex-valued, delocalized and concentrated on a central band that shrinks with system size.  The ring (a) has one pair of outermost unstable states which branches out from zero at $V_0=$ localize at a barrier edge, also present in the cut ring (b). Additionally, the cut ring has a pair of constant purely imaginary modes at around $\pm 2$ corresponding to instability localized at the system edge. (c) For box boundary, instabilities at the edge of the barrier (small loops) and system (large loops) vary periodically with $V_0$ as the $\rho'$ at the relevant edges varies with $V_0$.  The sloping curve in (b) is an envelope for the peaks of the smaller loops in (c).}
\label{Fig15_stability_modes}
\end{figure*}

The ${\rm Im}(\omega)$ are plotted in Fig.~\ref{Fig15_stability_modes} for the three cases.  We comment on the common features before we examine the differences. The bulk of the Bogoliubov modes are complex, as in they have a real and an imaginary part, and they correspond to a narrow band about zero, apart from a few exceptions we will identify below. These modes are delocalized and span the system and more importantly they go to zero as the system size is increased. As such, we can reasonably conclude that in the infinite limit these modes will not be a source of instability.

The most relevant modes are the purely imaginary ones and as the strength of the potential $V_0$ varies, they have conspicuously the largest absolute value and are separated from the band of complex eigenvalues. We find that with a few exceptions, there are always at most one or two conjugate pairs of such purely imaginary eigenvalues. These modes mark significant instabilities because they persist with increasing system size.

In the case of a ring, shown in Fig.~\ref{Fig15_stability_modes}(a) when $V_0=0$, ${\rm Im}(\omega)\simeq 0$ there are no instabilities. For $V\neq 0$ apart from the complex band, there are a series of splitting branches that increase in magnitude with stronger $V_0$. These mark the purely imaginary modes: For wells, $V_0<0$ there is only one such pair while for a barrier $V_0>0$ and there are one or two pairs. The instabilities localized at the potential edge correspond to the sole imaginary one or, if a pair, the one of larger $|{\rm Im}(\omega)|$ with the smaller one being delocalized like the complex modes. These last are absent with open boundary conditions, and hence we suspect they are due to the periodic boundary conditions and the fact that a non-zero rotation $\Omega$ is necessary to find the stationary solutions for each specific potential strength.  Furthermore, the jagged undulation on the trendlines on those purely imaginary modes are because of slight adjustments in potential width as the strength is changed to ensure identical solutions on either side, which is necessary for symmetric potential in a ring. As we increase the strength, the instability steadily increases due to the progressively sharper variation of the  mean density induced by the potential.

Comparing the ring with a cut ring in Fig.~\ref{Fig15_stability_modes}(b) confirms the points made above. To match the phase at the two edges use the same solution as for a ring, including adjusting $\Omega$, now an acceleration not a rotation for an open ring. Close to the center there is only one pair of imaginary roots, which match the localized modes in the ring, while the delocalized branches are absent. Additionally, there is a pair of purely imaginary modes $|{\rm Im}(\omega)|\simeq 2$, separated from the rest of the modes. These mark instability localized at the boundary. The value is constant with changing $V_0$ because we keep the endpoints at the same density and phase.

In the general case, Fig.~\ref{Fig15_stability_modes}(c), the value of the solutions at the ends of the system are not necessarily equal, and they vary with the potential strength. This creates an interesting periodic variation of the ${\rm Im}(\omega)$ of the mode that corresponds to the boundary instability as the density at the system edges change. These are manifest as the large loops seen in the figure. There is similar variation for the modes corresponding to the instability at the edge of the potential, seen as smaller loops. The trace of those modes in panel (b) form an envelope for these inner loops confirming their similar origin. We should stress that both set of loops are separated by regimes where those purely imaginary modes appear to be suppressed, these are regions where the derivative at the edges match the trend of change of the mean density across the edge of the potential or the system as a whole, in agreement with our main observation about the instabilities mentioned at the beginning of the section.

The instabilities at the system edges will be irrelevant in the infinite limit, but those at the barrier edges will persist. Even with finite systems, we conclude that those instabilities can be suppressed by adjusting the boundary to have density derivatives at the edges following the same trend as the change in the mean density. The sharpness of the transition accentuates the instability and will be softened for smoother potentials.  Here we considered instabilities for bounded oscillating solutions in all the potential regimes.  This limits the range of potential strengths $V_0$ since with values even slightly beyond what is shown in Fig.~\ref{Fig15_stability_modes} lead to unbounded solutions.

\begin{figure*}[t]
\centering
\includegraphics[width=\textwidth]{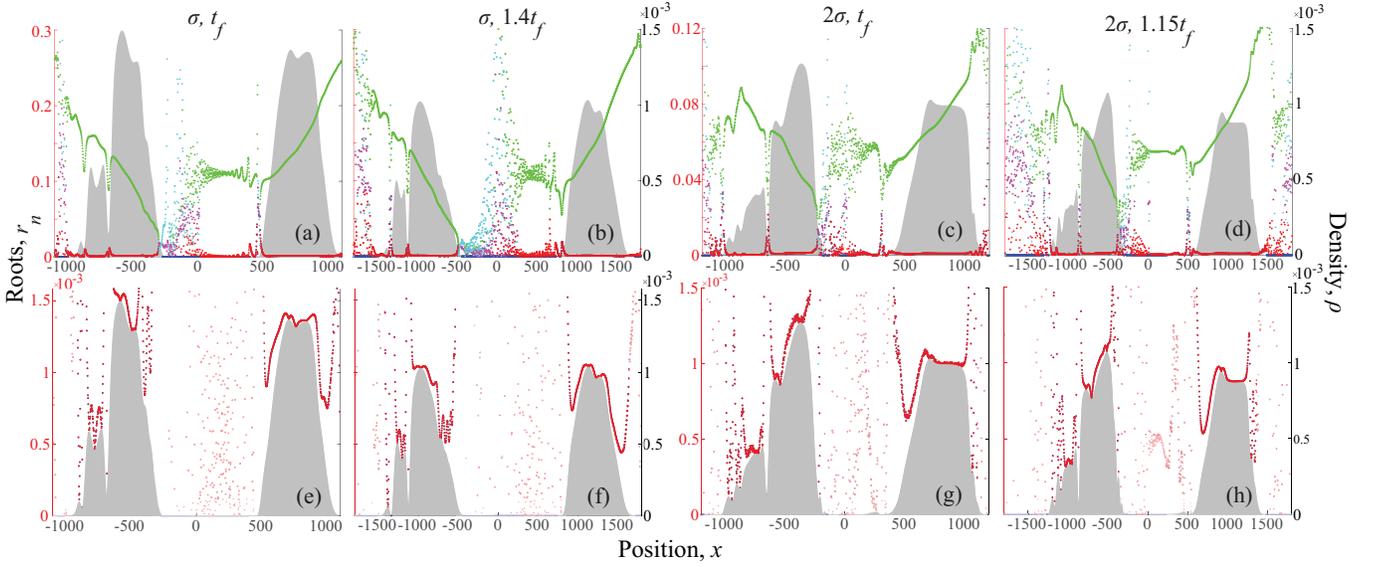}
\caption{Roots associated with stationary solutions are computed for the density and the phase values across the snapshot of a wavepacket (shaded) scattered by a barrier localized at the origin. In the upper panels all the roots are plotted as points: Real roots $\{r_1,r_2,r_3\}\rightarrow \{blue, red, green\}$, and for a complex root pair $\{r_1,{\rm Re}\{r_2,r_3\}, {\rm Im}\{r_2,r_3\}\}\rightarrow \{blue,magenta,cyan\}$. Four setups are shown; two with packet width $\sigma=80$ at (a,b) some time $t_f$ it takes for all of the density to scatter off and through the barrier and 1.4 times that time, and (c,d) two with packet width $2\sigma$ at $t_f$ and $1.15 t_f$ . The complex roots appear primarily in regimes of negligible density and hence are not relevant. The bottom panels are counterparts of the upper row, but zooming into the value of $r_1$ and $r_2$ showing that $r_1=r_2$ (hence only red dots are seen), which implies plane wave solutions.}
\label{Fig16_scattering}
\end{figure*}

\section{Applications to Scattering}

We finally address the question of the relevance of the stationary states to scattering by barrier potentials. Towards this end, we scatter a Gaussian wavepacket
\bn \psi=\frac{1}{\sqrt{\sigma\sqrt{\pi}}}e^{-(x-x_0)^2/2\sigma^2} e^{ikx}\en
that is launched towards a rectangular barrier and then examine the reflected and transmitted fractions after the scattering is complete, as has been utilized before in studying nonlinear scattering problems \cite{Das-PRA-R-Shift,Castin-nonlocal-superposition}. We use barrier of width $20$ and strength $V_0=0.1$ and wavepacket with velocity $k=0.5$. We show snapshots of the scattered wavepackets in gray filled silhouettes in Fig.~\ref{Fig16_scattering}.

For the linear Schr\"odinger equation, this method is very effective, with the results of the scattering matching analytical values for transmission and reflection \cite{Das-PRA-R-Shift}.  However, for the nonlinear problem it is more complicated. The lack of superposition principle makes it rather meaningless to define transmission and reflection amplitudes based on any stationary solution, although such attempts were made in a recent work \cite{A0}. Furthermore, the nonlinearity is proportional to the local density which will vary for a finite width wavepacket that would be typical in experiments, and hence the intrinsic non-uniformity of the packet poses a challenge.  Specifically, to address the latter issue,  we illustrate consistency for our method with two different packet widths, $\sigma=80$ and $160$, to assess how the features observed would behave in the limit of an infinite system. For meaningful comparison, we normalize the packets to unity and adjust the nonlinear strength $g$ to keep $g|\psi(x=x_0)|^2$ the same so that the nonlinearity at the peak density match for the wavepackets different width. Since the peak density scales as $\sigma^{-1}$ we use $g=2$, and $g=4$ respectively.

In order for our stationary solutions to be relevant for the scattering of finite wavepackets, we need to check whether the scattered packets can be represented with a finite set of those solutions. Since the Jacobi elliptic functions form an overcomplete basis, we must take a different approach than deconstructing them into Fourier components \cite{Handbook_elliptic}. Instead, we calculate directly the physical parameters $\alpha$, $\beta$, and $\mu$ at each point of the scattered wave packet at any specific instant of time. We do this by numerically computing the $\rho,\rho',\rho''$ and $\phi'$ at each point, using Eqs.~(\ref{eq1.2-1}),(\ref{eq1.2-2})and (\ref{eq.hd.3}).  Then using Eq.(\ref{parameters}), we can find the corresponding three roots $\{r_1,r_2,r_3\}$ at every point of the scattered packet.

We plot those roots in Fig.~\ref{Fig16_scattering}, superimposed on the scattered wave packets, as colored dots; with the three real roots in ascending order $\{r_1,r_2,r_3\}\rightarrow \{blue, red, green\}$, and for a pair of complex roots the real root is in $\{r_1,{\rm Re}\{r_2,r_3\}, {\rm Im}\{r_2,r_3\}\}\rightarrow \{blue,magenta,cyan\}$.  We show all the roots in the upper panels (a-d) of the figure, for two different widths and for each two different instants of time after scattering. What stands out in these panels is that the complex roots appear predominantly in the regimes of negligible density $\rho\simeq 0$ which means that only the real solutions are relevant.

The lower panels (e-h) zoom into the real solutions $r_1,r_2$ with $r_3$ off the scale, keeping the density profile at the same scale as in the upper panels. The panels illustrate a key feature, the two roots are equal $r_1=r_2$, and hence points of only one color is visible. Furthermore, since the real roots represent values of the density, our procedure is validated by the fact the roots follow the shape of the density of the wavepackets, where the density is significant, and moderately uniform. It is not surprising the solutions match plane waves, since even for linear scattering regime, density oscillations only arise due to interference between incident and reflected components, which is a consequence of a steady flow of an infinite stream.

As the scattered packets spread with time density of the plane waves will continue to decrease. However, as seen in Fig.~\ref{Fig16_scattering}, the roots we find continue to match the density profile regardless of the packet width. As the packet gets wider, there is a broader region of uniform density indicating a narrow range of solutions that can describe the bulk of the scattered packets.

We can therefore conclude that with sufficiently wide packets and with steady stream flow like in a nonlinear waveguide, the stationary solutions we derive can provide an adequate description of the scattering process, similarly to linear scattering. This method of analyzing a nonlinear scattering problem is more straightforward than previous methods utilized to study the same problem, which are heavily dependent on splitting solitons into free and trapped portions \cite{B3}, or using split-step Fourier methods with a hyperbolic function as an initial guess to solve the Gross-Pitaevskii equation \cite{B5}. This method of understanding scattering problems may shed light on BEC flow through a penetrable barrier \cite{BEC-flow-Atherton} or BEC behavior with impurities of either a single Gaussian defect or correlated disorder \cite{Hulet-BEC-transport}. The full landscape of solutions we have determined here can become relevant in situations where there is incidence from both directions and there is overlap between the scattered packets leading to density oscillations.

\section{Conclusions and Outlook}

We have determined the full landscape of stationary solutions of the quadratic nonlinear Schr\"odinger equation in the presence of a rectangular potential with different boundary conditions. As a preliminary, we determined the solutions at a potential step. A significant outcome is the inclusion of a class of solutions unbounded at infinity that have been generally left out in prior works, but are certainly relevant for finite width barriers as shown here.  We also find a simple unified expression in terms of a Jacobi elliptic function that describes the full spectrum of solutions, including those unbounded at infinity. In these regards, allowing for a complex phase shift is crucial, and that phase shift comes with some significant constraints required for physical solutions, both of which were overlooked previously.

Here we developed an approach based on using the roots and intersections of curves in phase space, that provides an intuitive way to construct and understand the solutions and the impact of boundary conditions. This method has allowed us to determine physical solutions and discard non-physical ones simply based on the location and features of those curves and their intersections. This method can find utility in understanding nonlinear systems and scattering problems with more generalized potentials and higher dimensionality.

The broad utility of the analytical solutions is established by our comparison with numerically computed solutions for a smooth barrier.  The close qualitative agreement of both the density and the phase show that the results obtained here would be relevant in experiments where smooth potentials are used, as in stirring of a BEC by sharply focused laser beam which would have a Gaussian profile similar to the one we used in our comparison.

We did a stability analysis of our solutions and found that persistent instabilities appear at the edge of the sharp boundaries either of the potential or the system as a whole. Specifically, the unstable modes localize at edges where the density derivative is correlates inversely to the actual trend of the density across the boundary.

On the motivating question of how stationary solutions are relevant in nonlinear scattering problems, considering the breakdown of the quantum superposition principle, we provide a definitive answer based on direct comparison with a numerically simulated scattering of a wavepacket. Instead of simply identifying density ratios as a measure of transmission, which has meaning only in the linear problem and only in the steady state, we take a practical approach, where we find the stationary solutions at each point of the scattered wave.  The close agreement of our solutions with the density profile, along with the finite range of such solutions needed at any instant in time, shows that the dynamical  nonlinear scattering can indeed be described with stationary solutions.  We expect that the full landscape of our stationary solutions can be used in future works to construct descriptions of more complex scattering scenarios that will included multiple input and output streams.

\begin{acknowledgments} We gratefully acknowledge the support of the NSF under Grant No. PHY-2011767. \end{acknowledgments}

\appendix

\section{Method for Analytical Solutions} \label{SecAppA:Methods}
To generate the analytical solutions for step, barrier, and well potentials, we utilized \emph{Mathematica}. The input parameters are the nonlinearity $g$, the potential height $V_0$, the barrier width $w$, the value of the density at the barrier $\rho_0$, and the three roots of $f_L$ satisfying the constraints discussed in Sec.~\ref{SecIII:Constraints}. Generally, when we want an oscillating solutions outside of the potential region, the value of $\rho_0$ is constrained to be between the two roots defining the loop in the $L$ region. In exceptional cases for a positive nonlinearity $g$ where asymptotic or decay solutions are permitted in the regions outside of the potential, $\rho_0$ is a value on the wing. Using the input parameters, Eq.~(\ref{eq.jacobidens}) is solved to determine $\rho_L(x)$, with the phase shift $x_{0L}$ determined by $\rho_0$ and a choice of the sign of the derivative. Using $f_L(\rho)$ as given in Eq.~(\ref{eq_f}), we find $f_P(\rho)$ and solve for the center roots. Knowing the value of the density at the edge of the potential, given by $\rho_0$, we then solve for $x_{0P}$ by inverting Eq.(\ref{eq.jacobidens}). We now have all the parameters needed to construct $\rho_P(x)$. A piecewise combination of the densities in both regions produces the solution for a step potential.

We follow similar steps to solve for the solution past the right barrier edge, $\rho_R(x)$, for a barrier or well potential. To find $f_R$, we use
\bn f_R=f_P-8V_0\rho(\rho-\rho_w)=f_L+8V_0\rho(\rho_w-\rho_0) \en
where $\rho_w=\rho_P(x=w)$. Using the value of the density at the right boundary $\rho_w$ to find $x_{0R}$ and the roots of $f_R$, we have all of the information necessary to construct $\rho_R(x)$. A piecewise combination of the densities in all three regions produces the full solution for a barrier or well potential.

Based on Eq.~(\ref{eq1.2-1}), we see that since the density is continuous, the phase will also be continuous. The phase may be solved for directly by substituting the solutions for the density in the appropriate regions. On a ring, the periodic boundary condition creates another constraint; an additional linear term involving rotation ensures that the phase is continuous and differentiable at all points on the ring.

\section{Jacobi Elliptic Functions with Complex arguments} \label{SecAppB:Elliptic}

In this appendix, we will derive Eq.~(\ref{eq.dens2}) from Eq.~(\ref{eq.jacobidens}) with the only assumption that $g<0$. We utilize Jacobi elliptic identities as given in \cite{Handbook_elliptic}. Consider a ${\rm {\rm sn}}$ function with a complex argument
\begin{multline} \label{eq.complexsn}
{\rm sn}(u+iv,m) = \frac{1}{1-{\rm sn}^2(v,m'){\rm dn}^2(u,m)}...\\
...\times[{\rm sn}(u,m){\rm dn}(v,m')...\\
...+i{\rm cn}(u,m){\rm dn}(u,m){\rm sn}(v,m'){\rm cn}(v,m')]
\end{multline}
Since $g(r_3-r_1)$ is negative, the only real component is that of $x_0$. The imaginary part must vanish else the density will be imaginary or negative, and we can ensure that provided the real part of $x_0$ is an elliptic integral of the first kind,
\bn {\rm Re}(x_{0})=K\left(\frac{r_2-r_1}{r_3-r_1}\right) \en
This ensures that only arguments with $v$ remain. When our argument $u$ is given by an elliptic integral of the modulus used, we obtain the following values:
\bn {\rm sn}(K(m),m)=1 \n \\ {\rm cn}(K(m),m)=0 \n \\ {\rm dn}(K(m),m)=\sqrt{m'} \label{eq.jacobiK} \en

These limiting values, combined with the identity
\bn {\rm dn}^2(u,m)+m{\rm sn}^2(u,m)=1 \label{dnsn}\en
reduce our expression to
\bn {\rm sn}(K(m)+iv,m) &=&\frac{{\rm dn}(v,m')}{1-m'{\rm sn}^2(v,m')} \nonumber \\
               &=&\frac{1}{{\rm dn}(v,m')} \en
Squaring both sides and again using Eq.~({\ref{dnsn}) we obtain
\bn {\rm sn}^2(K(m)+iv,m)&=&\frac{m'{\rm sn}^2(v,m')+{\rm dn}^2(v,m')}{{\rm dn}^2(v,m')} \n \\
                &=&1+m'{\rm sd}^2(v,m') \en
Inserting this back into Eq.~(\ref{eq.jacobidens}), we obtain
\bn \rho &=&r_2+\frac{(r_2-r_1)(r_3-r_2)}{r_3-r_1}\\
&& \times{\rm sd}^2\left(\sqrt{|g|(r_3-r_1)}x+{\rm Im}(x_{0}),\frac{r_3-r_2}{r_3-r_1}\right)\n
\en
Then we use the relation
\bn {\rm cn}(u+K,m)&=&-\sqrt{m'}{\rm sd}(u,m)\\\n \en
and redfine $x_0=K(m)+{\rm Im}(x_0)$, to write the density as
\bn \rho=r_2+(r_3-r_2){\rm cn}^2\left(\sqrt{|g|(r_3-r_1)}x+x_0,\frac{r_3-r_2}{r_3-r_1}\right) \n \\
 \en%
Finally, using the identity
\bn {\rm sn}^2(u,m)+{\rm cn}^2(u,m)=1 \en
we arrive at the form used in Eq.~(\ref{eq.dens2})
\bn \label{eq.altdens} \rho=r_3+(r_2-r_3){\rm sn}^2\left(\sqrt{|g|(r_3-r_1)}x+x_0,\frac{r_3-r_2}{r_3-r_1}\right) \n \\
\en

This is a far more intuitive equation for the density, as the argument is real and elliptic modulus is between zero and one. From this, we see that $0\leq {\rm sn}^2\leq1$, and the density oscillates between $r_2$ and $r_3$. We could also have derived this same equation through a symmetry argument. Looking at Eq.~(\ref{eq.jacobidens}), we make the argument that flipping the sign of $g$ is simply like reversing the behaviors of the roots, so that $r_1$ takes on the role of $r_3$ and vice versa. By simply swapping $r_1$ and $r_3$ in Eq.~(\ref{eq.jacobidens}), we recover Eq.~(\ref{eq.altdens}).

\vfill


\begin{thebibliography}{63}%
\makeatletter
\providecommand \@ifxundefined [1]{%
 \@ifx{#1\undefined}
}%
\providecommand \@ifnum [1]{%
 \ifnum #1\expandafter \@firstoftwo
 \else \expandafter \@secondoftwo
 \fi
}%
\providecommand \@ifx [1]{%
 \ifx #1\expandafter \@firstoftwo
 \else \expandafter \@secondoftwo
 \fi
}%
\providecommand \natexlab [1]{#1}%
\providecommand \enquote  [1]{``#1''}%
\providecommand \bibnamefont  [1]{#1}%
\providecommand \bibfnamefont [1]{#1}%
\providecommand \citenamefont [1]{#1}%
\providecommand \href@noop [0]{\@secondoftwo}%
\providecommand \href [0]{\begingroup \@sanitize@url \@href}%
\providecommand \@href[1]{\@@startlink{#1}\@@href}%
\providecommand \@@href[1]{\endgroup#1\@@endlink}%
\providecommand \@sanitize@url [0]{\catcode `\\12\catcode `\$12\catcode
  `\&12\catcode `\#12\catcode `\^12\catcode `\_12\catcode `\%12\relax}%
\providecommand \@@startlink[1]{}%
\providecommand \@@endlink[0]{}%
\providecommand \url  [0]{\begingroup\@sanitize@url \@url }%
\providecommand \@url [1]{\endgroup\@href {#1}{\urlprefix }}%
\providecommand \urlprefix  [0]{URL }%
\providecommand \Eprint [0]{\href }%
\providecommand \doibase [0]{https://doi.org/}%
\providecommand \selectlanguage [0]{\@gobble}%
\providecommand \bibinfo  [0]{\@secondoftwo}%
\providecommand \bibfield  [0]{\@secondoftwo}%
\providecommand \translation [1]{[#1]}%
\providecommand \BibitemOpen [0]{}%
\providecommand \bibitemStop [0]{}%
\providecommand \bibitemNoStop [0]{.\EOS\space}%
\providecommand \EOS [0]{\spacefactor3000\relax}%
\providecommand \BibitemShut  [1]{\csname bibitem#1\endcsname}%
\let\auto@bib@innerbib\@empty
\bibitem [{\citenamefont {Mahan}(2012)}]{mahan}%
  \BibitemOpen
  \bibfield  {author} {\bibinfo {author} {\bibfnamefont {G.~D.}\ \bibnamefont
  {Mahan}},\ }\href@noop {} {\emph {\bibinfo {title} {Many-Particle
  Physics}}},\ \bibinfo {edition} {2nd}\ ed.\ (\bibinfo  {publisher}
  {Springer},\ \bibinfo {address} {USA},\ \bibinfo {year} {2012})\BibitemShut
  {NoStop}%
\bibitem [{\citenamefont {Liboff}(2002)}]{liboff}%
  \BibitemOpen
  \bibfield  {author} {\bibinfo {author} {\bibfnamefont {R.~L.}\ \bibnamefont
  {Liboff}},\ }\href@noop {} {\emph {\bibinfo {title} {Introductory Quantum
  Mechanics}}},\ \bibinfo {edition} {4th}\ ed.\ (\bibinfo  {publisher}
  {Addison-Wesley},\ \bibinfo {address} {USA},\ \bibinfo {year}
  {2002})\BibitemShut {NoStop}%
\bibitem [{\citenamefont {Dalfovo}\ \emph {et~al.}(1999)\citenamefont
  {Dalfovo}, \citenamefont {Giorgini}, \citenamefont {Pitaevskii},\ and\
  \citenamefont {Stringari}}]{RMP-Sringari-1999}%
  \BibitemOpen
  \bibfield  {author} {\bibinfo {author} {\bibfnamefont {F.}~\bibnamefont
  {Dalfovo}}, \bibinfo {author} {\bibfnamefont {S.}~\bibnamefont {Giorgini}},
  \bibinfo {author} {\bibfnamefont {L.~P.}\ \bibnamefont {Pitaevskii}},\ and\
  \bibinfo {author} {\bibfnamefont {S.}~\bibnamefont {Stringari}},\ }\bibfield
  {title} {\bibinfo {title} {Theory of Bose-Einstein condensation in trapped
  gases},\ }\href {https://doi.org/10.1103/RevModPhys.71.463} {\bibfield
  {journal} {\bibinfo  {journal} {Rev. Mod. Phys.}\ }\textbf {\bibinfo {volume}
  {71}},\ \bibinfo {pages} {463} (\bibinfo {year} {1999})}\BibitemShut
  {NoStop}%
\bibitem [{\citenamefont {Chiao}\ \emph {et~al.}(1964)\citenamefont {Chiao},
  \citenamefont {Garmire},\ and\ \citenamefont {Townes}}]{NLSE-Chiao-Townes}%
  \BibitemOpen
  \bibfield  {author} {\bibinfo {author} {\bibfnamefont {R.~Y.}\ \bibnamefont
  {Chiao}}, \bibinfo {author} {\bibfnamefont {E.}~\bibnamefont {Garmire}},\
  and\ \bibinfo {author} {\bibfnamefont {C.~H.}\ \bibnamefont {Townes}},\
  }\bibfield  {title} {\bibinfo {title} {Self-trapping of optical beams},\
  }\href {https://doi.org/10.1103/PhysRevLett.13.479} {\bibfield  {journal}
  {\bibinfo  {journal} {Phys. Rev. Lett.}\ }\textbf {\bibinfo {volume} {13}},\
  \bibinfo {pages} {479} (\bibinfo {year} {1964})}\BibitemShut {NoStop}%
\bibitem [{\citenamefont {Boyd}\ and\ \citenamefont {Prato}(2008)}]{Boyd}%
  \BibitemOpen
  \bibfield  {author} {\bibinfo {author} {\bibfnamefont {R.~W.}\ \bibnamefont
  {Boyd}}\ and\ \bibinfo {author} {\bibfnamefont {D.}~\bibnamefont {Prato}},\
  }\href@noop {} {\emph {\bibinfo {title} {Nonlinear Optics}}},\ \bibinfo
  {edition} {3rd}\ ed.\ (\bibinfo  {publisher} {Academic Press},\ \bibinfo
  {address} {Burlington, MA},\ \bibinfo {year} {2008})\BibitemShut {NoStop}%
\bibitem [{\citenamefont {Agarwal}(2001)}]{G.P.Agarwal}%
  \BibitemOpen
  \bibfield  {author} {\bibinfo {author} {\bibfnamefont {G.~P.}\ \bibnamefont
  {Agarwal}},\ }\href@noop {} {\emph {\bibinfo {title} {Nonlinear Fiber
  Optics}}}\ (\bibinfo  {publisher} {Academic press},\ \bibinfo {address} {San
  Diego, CA},\ \bibinfo {year} {2001})\BibitemShut {NoStop}%
\bibitem [{\citenamefont {Pitaevskii}\ and\ \citenamefont
  {Stringar}(2003)}]{Pitaevskii-Stringari}%
  \BibitemOpen
  \bibfield  {author} {\bibinfo {author} {\bibfnamefont {L.}~\bibnamefont
  {Pitaevskii}}\ and\ \bibinfo {author} {\bibfnamefont {S.}~\bibnamefont
  {Stringar}},\ }\href@noop {} {\emph {\bibinfo {title} {Bose Einstein
  Condensation}}},\ \bibinfo {edition} {1st}\ ed.\ (\bibinfo  {publisher}
  {Clarendaon Press},\ \bibinfo {year} {2003})\BibitemShut {NoStop}%
\bibitem [{\citenamefont {Pethick}\ and\ \citenamefont
  {Smith}(2008)}]{Pethick-Smith}%
  \BibitemOpen
  \bibfield  {author} {\bibinfo {author} {\bibfnamefont {C.~J.}\ \bibnamefont
  {Pethick}}\ and\ \bibinfo {author} {\bibfnamefont {H.}~\bibnamefont
  {Smith}},\ }\href@noop {} {\emph {\bibinfo {title} {Bose Einstein
  Condensation in Dilute Gases}}},\ \bibinfo {edition} {2nd}\ ed.\ (\bibinfo
  {publisher} {Cambridge University Press},\ \bibinfo {address} {UK},\ \bibinfo
  {year} {2008})\BibitemShut {NoStop}%
\bibitem [{\citenamefont {Kevrekidis}\ \emph {et~al.}(2008)\citenamefont
  {Kevrekidis}, \citenamefont {Frantzeskakis},\ and\ \citenamefont
  {Carretero-Gonz\'alez}}]{Kevrekidis-Carretero}%
  \BibitemOpen
  \bibfield  {author} {\bibinfo {author} {\bibfnamefont {P.~G.}\ \bibnamefont
  {Kevrekidis}}, \bibinfo {author} {\bibfnamefont {D.~J.}\ \bibnamefont
  {Frantzeskakis}},\ and\ \bibinfo {author} {\bibfnamefont {R.}~\bibnamefont
  {Carretero-Gonz\'alez}},\ }\href@noop {} {\emph {\bibinfo {title} {Emergent
  Nonlinear Phenomena in Bose Einstein Condensates}}},\ \bibinfo {edition}
  {4th}\ ed.\ (\bibinfo  {publisher} {Springer},\ \bibinfo {address} {Berlin},\
  \bibinfo {year} {2008})\BibitemShut {NoStop}%
\bibitem [{\citenamefont {Al-Khawaja}\ and\ \citenamefont
  {Al~Sakkaf}(2019)}]{Khawaja-NLSE}%
  \BibitemOpen
  \bibfield  {author} {\bibinfo {author} {\bibfnamefont {U.}~\bibnamefont
  {Al-Khawaja}}\ and\ \bibinfo {author} {\bibfnamefont {L.}~\bibnamefont
  {Al~Sakkaf}},\ }\href@noop {} {\emph {\bibinfo {title} {Handbook of Exact
  Solutions to the Nonlinear Schr\"odinger Equations}}}\ (\bibinfo  {publisher}
  {IOP Publishing},\ \bibinfo {address} {Bristol, UK},\ \bibinfo {year}
  {2019})\BibitemShut {NoStop}%
\bibitem [{\citenamefont {Fibich}(2015)}]{Fibich-NLSE}%
  \BibitemOpen
  \bibfield  {author} {\bibinfo {author} {\bibfnamefont {G.}~\bibnamefont
  {Fibich}},\ }\href@noop {} {\emph {\bibinfo {title} {The Nonlinear
  Schr\"odinger Equation: Singular Solutions and Optical Collapse}}}\ (\bibinfo
   {publisher} {Springer},\ \bibinfo {address} {Switzerland},\ \bibinfo {year}
  {2015})\BibitemShut {NoStop}%
\bibitem [{\citenamefont {{Zakharov}}\ and\ \citenamefont
  {{Shabat}}(1972)}]{Zakharov-JETP}%
  \BibitemOpen
  \bibfield  {author} {\bibinfo {author} {\bibfnamefont {V.~E.}\ \bibnamefont
  {{Zakharov}}}\ and\ \bibinfo {author} {\bibfnamefont {A.~B.}\ \bibnamefont
  {{Shabat}}},\ }\bibfield  {title} {\bibinfo {title} {{Exact Theory of
  Two-dimensional Self-focusing and One-dimensional Self-modulation of Waves in
  Nonlinear Media}},\ }\href@noop {} {\bibfield  {journal} {\bibinfo  {journal}
  {Soviet Journal of Experimental and Theoretical Physics}\ }\textbf {\bibinfo
  {volume} {34}},\ \bibinfo {pages} {62} (\bibinfo {year} {1972})}\BibitemShut
  {NoStop}%
\bibitem [{\citenamefont {Carr}\ \emph
  {et~al.}(2000{\natexlab{a}})\citenamefont {Carr}, \citenamefont {Clark},\
  and\ \citenamefont {Reinhardt}}]{A1}%
  \BibitemOpen
  \bibfield  {author} {\bibinfo {author} {\bibfnamefont {L.~D.}\ \bibnamefont
  {Carr}}, \bibinfo {author} {\bibfnamefont {C.~W.}\ \bibnamefont {Clark}},\
  and\ \bibinfo {author} {\bibfnamefont {W.~P.}\ \bibnamefont {Reinhardt}},\
  }\bibfield  {title} {\bibinfo {title} {Stationary solutions of the
  one-dimensional nonlinear Schrodinger equation. I. case of repulsive
  nonlinearity},\ }\href {https://doi.org/10.1103/PhysRevA.62.063610}
  {\bibfield  {journal} {\bibinfo  {journal} {Phys. Rev. A}\ }\textbf {\bibinfo
  {volume} {62}},\ \bibinfo {pages} {063610} (\bibinfo {year}
  {2000}{\natexlab{a}})}\BibitemShut {NoStop}%
\bibitem [{\citenamefont {Carr}\ \emph
  {et~al.}(2000{\natexlab{b}})\citenamefont {Carr}, \citenamefont {Clark},\
  and\ \citenamefont {Reinhardt}}]{A2}%
  \BibitemOpen
  \bibfield  {author} {\bibinfo {author} {\bibfnamefont {L.~D.}\ \bibnamefont
  {Carr}}, \bibinfo {author} {\bibfnamefont {C.~W.}\ \bibnamefont {Clark}},\
  and\ \bibinfo {author} {\bibfnamefont {W.~P.}\ \bibnamefont {Reinhardt}},\
  }\bibfield  {title} {\bibinfo {title} {Stationary solutions of the
  one-dimensional nonlinear Schrodinger equation. II. case of attractive
  nonlinearity},\ }\href {https://doi.org/10.1103/PhysRevA.62.063611}
  {\bibfield  {journal} {\bibinfo  {journal} {Phys. Rev. A}\ }\textbf {\bibinfo
  {volume} {62}},\ \bibinfo {pages} {063611} (\bibinfo {year}
  {2000}{\natexlab{b}})}\BibitemShut {NoStop}%
\bibitem [{\citenamefont {Dror}\ and\ \citenamefont {Malomed}(2015)}]{B13}%
  \BibitemOpen
  \bibfield  {author} {\bibinfo {author} {\bibfnamefont {N.}~\bibnamefont
  {Dror}}\ and\ \bibinfo {author} {\bibfnamefont {B.~A.}\ \bibnamefont
  {Malomed}},\ }\bibfield  {title} {\bibinfo {title} {Solitons and vortices in
  nonlinear potential wells},\ }\href@noop {} {\bibfield  {journal} {\bibinfo
  {journal} {Journal of Optics}\ }\textbf {\bibinfo {volume} {18}},\ \bibinfo
  {pages} {014003} (\bibinfo {year} {2015})}\BibitemShut {NoStop}%
\bibitem [{\citenamefont {Serkin}\ and\ \citenamefont {Hasegawa}(2000)}]{B34}%
  \BibitemOpen
  \bibfield  {author} {\bibinfo {author} {\bibfnamefont {V.~N.}\ \bibnamefont
  {Serkin}}\ and\ \bibinfo {author} {\bibfnamefont {A.}~\bibnamefont
  {Hasegawa}},\ }\bibfield  {title} {\bibinfo {title} {Novel soliton solutions
  of the nonlinear Schr\"odinger equation model},\ }\href
  {https://doi.org/10.1103/PhysRevLett.85.4502} {\bibfield  {journal} {\bibinfo
   {journal} {Phys. Rev. Lett.}\ }\textbf {\bibinfo {volume} {85}},\ \bibinfo
  {pages} {4502} (\bibinfo {year} {2000})}\BibitemShut {NoStop}%
\bibitem [{\citenamefont {Seaman}\ \emph
  {et~al.}(2005{\natexlab{a}})\citenamefont {Seaman}, \citenamefont {Carr},\
  and\ \citenamefont {Holland}}]{A4}%
  \BibitemOpen
  \bibfield  {author} {\bibinfo {author} {\bibfnamefont {B.~T.}\ \bibnamefont
  {Seaman}}, \bibinfo {author} {\bibfnamefont {L.~D.}\ \bibnamefont {Carr}},\
  and\ \bibinfo {author} {\bibfnamefont {M.~J.}\ \bibnamefont {Holland}},\
  }\bibfield  {title} {\bibinfo {title} {Nonlinear band structure in
  Bose-Einstein condensates: Nonlinear Schr\"odinger equation with a
  Kronig-Penney potential},\ }\href
  {https://doi.org/10.1103/PhysRevA.71.033622} {\bibfield  {journal} {\bibinfo
  {journal} {Phys. Rev. A}\ }\textbf {\bibinfo {volume} {71}},\ \bibinfo
  {pages} {033622} (\bibinfo {year} {2005}{\natexlab{a}})}\BibitemShut
  {NoStop}%
\bibitem [{\citenamefont {Hu}\ \emph {et~al.}(2012)\citenamefont {Hu},
  \citenamefont {Ma}, \citenamefont {Lu}, \citenamefont {Zheng},\ and\
  \citenamefont {Hu}}]{B10}%
  \BibitemOpen
  \bibfield  {author} {\bibinfo {author} {\bibfnamefont {S.}~\bibnamefont
  {Hu}}, \bibinfo {author} {\bibfnamefont {X.}~\bibnamefont {Ma}}, \bibinfo
  {author} {\bibfnamefont {D.}~\bibnamefont {Lu}}, \bibinfo {author}
  {\bibfnamefont {Y.}~\bibnamefont {Zheng}},\ and\ \bibinfo {author}
  {\bibfnamefont {W.}~\bibnamefont {Hu}},\ }\bibfield  {title} {\bibinfo
  {title} {Defect solitons in parity-time-symmetric optical lattices with
  nonlocal nonlinearity},\ }\href {https://doi.org/10.1103/PhysRevA.85.043826}
  {\bibfield  {journal} {\bibinfo  {journal} {Phys. Rev. A}\ }\textbf {\bibinfo
  {volume} {85}},\ \bibinfo {pages} {043826} (\bibinfo {year}
  {2012})}\BibitemShut {NoStop}%
\bibitem [{\citenamefont {Zhang}\ \emph {et~al.}(2010)\citenamefont {Zhang},
  \citenamefont {Li}, \citenamefont {Meng}, \citenamefont {Wu},\ and\
  \citenamefont {Malomed}}]{B11}%
  \BibitemOpen
  \bibfield  {author} {\bibinfo {author} {\bibfnamefont {J.-F.}\ \bibnamefont
  {Zhang}}, \bibinfo {author} {\bibfnamefont {Y.-S.}\ \bibnamefont {Li}},
  \bibinfo {author} {\bibfnamefont {J.}~\bibnamefont {Meng}}, \bibinfo {author}
  {\bibfnamefont {L.}~\bibnamefont {Wu}},\ and\ \bibinfo {author}
  {\bibfnamefont {B.~A.}\ \bibnamefont {Malomed}},\ }\bibfield  {title}
  {\bibinfo {title} {Matter-wave solitons and finite-amplitude bloch waves in
  optical lattices with spatially modulated nonlinearity},\ }\href
  {https://doi.org/10.1103/PhysRevA.82.033614} {\bibfield  {journal} {\bibinfo
  {journal} {Phys. Rev. A}\ }\textbf {\bibinfo {volume} {82}},\ \bibinfo
  {pages} {033614} (\bibinfo {year} {2010})}\BibitemShut {NoStop}%
\bibitem [{\citenamefont {Alatas}(2011)}]{B15}%
  \BibitemOpen
  \bibfield  {author} {\bibinfo {author} {\bibfnamefont {H.}~\bibnamefont
  {Alatas}},\ }\bibfield  {title} {\bibinfo {title} {Dynamics of Jacobi's
  elliptic spatial waves in a nonlinear optical grating},\ }\href
  {https://doi.org/10.1103/PhysRevA.83.043830} {\bibfield  {journal} {\bibinfo
  {journal} {Phys. Rev. A}\ }\textbf {\bibinfo {volume} {83}},\ \bibinfo
  {pages} {043830} (\bibinfo {year} {2011})}\BibitemShut {NoStop}%
\bibitem [{\citenamefont {Petrovi\ifmmode~\acute{c}\else \'{c}\fi{}}\ \emph
  {et~al.}(2011)\citenamefont {Petrovi\ifmmode~\acute{c}\else \'{c}\fi{}},
  \citenamefont {Beli\ifmmode~\acute{c}\else \'{c}\fi{}},\ and\ \citenamefont
  {Zhong}}]{B16}%
  \BibitemOpen
  \bibfield  {author} {\bibinfo {author} {\bibfnamefont {N.~Z.}\ \bibnamefont
  {Petrovi\ifmmode~\acute{c}\else \'{c}\fi{}}}, \bibinfo {author}
  {\bibfnamefont {M.}~\bibnamefont {Beli\ifmmode~\acute{c}\else \'{c}\fi{}}},\
  and\ \bibinfo {author} {\bibfnamefont {W.-P.}\ \bibnamefont {Zhong}},\
  }\bibfield  {title} {\bibinfo {title} {Exact traveling-wave and
  spatiotemporal soliton solutions to the generalized ($3+1$)-dimensional
  Schr\"odinger equation with polynomial nonlinearity of arbitrary order},\
  }\href {https://doi.org/10.1103/PhysRevE.83.026604} {\bibfield  {journal}
  {\bibinfo  {journal} {Phys. Rev. E}\ }\textbf {\bibinfo {volume} {83}},\
  \bibinfo {pages} {026604} (\bibinfo {year} {2011})}\BibitemShut {NoStop}%
\bibitem [{\citenamefont {Hasegawa}\ and\ \citenamefont
  {Tappert}(1973{\natexlab{a}})}]{B21_1}%
  \BibitemOpen
  \bibfield  {author} {\bibinfo {author} {\bibfnamefont {A.}~\bibnamefont
  {Hasegawa}}\ and\ \bibinfo {author} {\bibfnamefont {F.}~\bibnamefont
  {Tappert}},\ }\bibfield  {title} {\bibinfo {title} {Transmission of
  stationary nonlinear optical pulses in dispersive dielectric fibers. I.
  anomalous dispersion},\ }\href {https://doi.org/10.1063/1.1654836} {\bibfield
   {journal} {\bibinfo  {journal} {Applied Physics Letters}\ }\textbf {\bibinfo
  {volume} {23}},\ \bibinfo {pages} {142} (\bibinfo {year}
  {1973}{\natexlab{a}})}\BibitemShut {NoStop}%
\bibitem [{\citenamefont {Hasegawa}\ and\ \citenamefont
  {Tappert}(1973{\natexlab{b}})}]{B21_2}%
  \BibitemOpen
  \bibfield  {author} {\bibinfo {author} {\bibfnamefont {A.}~\bibnamefont
  {Hasegawa}}\ and\ \bibinfo {author} {\bibfnamefont {F.}~\bibnamefont
  {Tappert}},\ }\bibfield  {title} {\bibinfo {title} {Transmission of
  stationary nonlinear optical pulses in dispersive dielectric fibers. II.
  normal dispersion},\ }\href {https://doi.org/10.1063/1.1654836} {\bibfield
  {journal} {\bibinfo  {journal} {Applied Physics Letters}\ }\textbf {\bibinfo
  {volume} {23}},\ \bibinfo {pages} {171} (\bibinfo {year}
  {1973}{\natexlab{b}})}\BibitemShut {NoStop}%
\bibitem [{\citenamefont {Bingzhen}\ and\ \citenamefont
  {Wenzheng}(1995)}]{B14}%
  \BibitemOpen
  \bibfield  {author} {\bibinfo {author} {\bibfnamefont {X.}~\bibnamefont
  {Bingzhen}}\ and\ \bibinfo {author} {\bibfnamefont {W.}~\bibnamefont
  {Wenzheng}},\ }\bibfield  {title} {\bibinfo {title} {Traveling-wave method
  for solving the modified nonlinear Schr\"odinger equation describing soliton
  propagation along optical fibers},\ }\href
  {https://doi.org/10.1103/PhysRevE.51.1493} {\bibfield  {journal} {\bibinfo
  {journal} {Phys. Rev. E}\ }\textbf {\bibinfo {volume} {51}},\ \bibinfo
  {pages} {1493} (\bibinfo {year} {1995})}\BibitemShut {NoStop}%
\bibitem [{\citenamefont {de~Oliveira}\ and\ \citenamefont
  {de~Moura}(1998)}]{B18}%
  \BibitemOpen
  \bibfield  {author} {\bibinfo {author} {\bibfnamefont {J.~R.}\ \bibnamefont
  {de~Oliveira}}\ and\ \bibinfo {author} {\bibfnamefont {M.~A.}\ \bibnamefont
  {de~Moura}},\ }\bibfield  {title} {\bibinfo {title} {Analytical solution for
  the modified nonlinear Schr\"odinger equation describing optical shock
  formation},\ }\href {https://doi.org/10.1103/PhysRevE.57.4751} {\bibfield
  {journal} {\bibinfo  {journal} {Phys. Rev. E}\ }\textbf {\bibinfo {volume}
  {57}},\ \bibinfo {pages} {4751} (\bibinfo {year} {1998})}\BibitemShut
  {NoStop}%
\bibitem [{\citenamefont {Hiro-Oka}\ and\ \citenamefont
  {Minakata}(1994)}]{B20}%
  \BibitemOpen
  \bibfield  {author} {\bibinfo {author} {\bibfnamefont {H.}~\bibnamefont
  {Hiro-Oka}}\ and\ \bibinfo {author} {\bibfnamefont {H.}~\bibnamefont
  {Minakata}},\ }\bibfield  {title} {\bibinfo {title} {Solitons in the
  nonlinear Schr\"odinger model and the collective ground state of a
  one-dimensional delta-function gas},\ }\href
  {https://doi.org/https://doi.org/10.1016/0375-9601(94)90153-8} {\bibfield
  {journal} {\bibinfo  {journal} {Physics Letters A}\ }\textbf {\bibinfo
  {volume} {195}},\ \bibinfo {pages} {204} (\bibinfo {year}
  {1994})}\BibitemShut {NoStop}%
\bibitem [{\citenamefont {Kruglov}\ and\ \citenamefont {Harvey}(2018)}]{B26}%
  \BibitemOpen
  \bibfield  {author} {\bibinfo {author} {\bibfnamefont {V.~I.}\ \bibnamefont
  {Kruglov}}\ and\ \bibinfo {author} {\bibfnamefont {J.~D.}\ \bibnamefont
  {Harvey}},\ }\bibfield  {title} {\bibinfo {title} {Solitary waves in optical
  fibers governed by higher-order dispersion},\ }\href
  {https://doi.org/10.1103/PhysRevA.98.063811} {\bibfield  {journal} {\bibinfo
  {journal} {Phys. Rev. A}\ }\textbf {\bibinfo {volume} {98}},\ \bibinfo
  {pages} {063811} (\bibinfo {year} {2018})}\BibitemShut {NoStop}%
\bibitem [{\citenamefont {Rohrmann}\ \emph {et~al.}(2013)\citenamefont
  {Rohrmann}, \citenamefont {Hause},\ and\ \citenamefont {Mitschke}}]{B27}%
  \BibitemOpen
  \bibfield  {author} {\bibinfo {author} {\bibfnamefont {P.}~\bibnamefont
  {Rohrmann}}, \bibinfo {author} {\bibfnamefont {A.}~\bibnamefont {Hause}},\
  and\ \bibinfo {author} {\bibfnamefont {F.}~\bibnamefont {Mitschke}},\
  }\bibfield  {title} {\bibinfo {title} {Two-soliton and three-soliton
  molecules in optical fibers},\ }\href
  {https://doi.org/10.1103/PhysRevA.87.043834} {\bibfield  {journal} {\bibinfo
  {journal} {Phys. Rev. A}\ }\textbf {\bibinfo {volume} {87}},\ \bibinfo
  {pages} {043834} (\bibinfo {year} {2013})}\BibitemShut {NoStop}%
\bibitem [{\citenamefont {Renninger}\ and\ \citenamefont {Wise}(2013)}]{B35}%
  \BibitemOpen
  \bibfield  {author} {\bibinfo {author} {\bibfnamefont {W.~H.}\ \bibnamefont
  {Renninger}}\ and\ \bibinfo {author} {\bibfnamefont {F.~W.}\ \bibnamefont
  {Wise}},\ }\bibfield  {title} {\bibinfo {title} {Optical solitons in
  graded-index multimode fibres},\ }\href {https://doi.org/10.1038/ncomms2739}
  {\bibfield  {journal} {\bibinfo  {journal} {Nature Communications}\ }\textbf
  {\bibinfo {volume} {4}},\ \bibinfo {pages} {1719} (\bibinfo {year}
  {2013})}\BibitemShut {NoStop}%
\bibitem [{\citenamefont {Rand}\ \emph {et~al.}(2007)\citenamefont {Rand},
  \citenamefont {Glesk}, \citenamefont {Br\`es}, \citenamefont {Nolan},
  \citenamefont {Chen}, \citenamefont {Koh}, \citenamefont {Fleischer},
  \citenamefont {Steiglitz},\ and\ \citenamefont {Prucnal}}]{B32}%
  \BibitemOpen
  \bibfield  {author} {\bibinfo {author} {\bibfnamefont {D.}~\bibnamefont
  {Rand}}, \bibinfo {author} {\bibfnamefont {I.}~\bibnamefont {Glesk}},
  \bibinfo {author} {\bibfnamefont {C.-S.}\ \bibnamefont {Br\`es}}, \bibinfo
  {author} {\bibfnamefont {D.~A.}\ \bibnamefont {Nolan}}, \bibinfo {author}
  {\bibfnamefont {X.}~\bibnamefont {Chen}}, \bibinfo {author} {\bibfnamefont
  {J.}~\bibnamefont {Koh}}, \bibinfo {author} {\bibfnamefont {J.~W.}\
  \bibnamefont {Fleischer}}, \bibinfo {author} {\bibfnamefont {K.}~\bibnamefont
  {Steiglitz}},\ and\ \bibinfo {author} {\bibfnamefont {P.~R.}\ \bibnamefont
  {Prucnal}},\ }\bibfield  {title} {\bibinfo {title} {Observation of temporal
  vector soliton propagation and collision in birefringent fiber},\ }\href
  {https://doi.org/10.1103/PhysRevLett.98.053902} {\bibfield  {journal}
  {\bibinfo  {journal} {Phys. Rev. Lett.}\ }\textbf {\bibinfo {volume} {98}},\
  \bibinfo {pages} {053902} (\bibinfo {year} {2007})}\BibitemShut {NoStop}%
\bibitem [{\citenamefont {Ernst}\ and\ \citenamefont {Brand}(2010)}]{B3}%
  \BibitemOpen
  \bibfield  {author} {\bibinfo {author} {\bibfnamefont {T.}~\bibnamefont
  {Ernst}}\ and\ \bibinfo {author} {\bibfnamefont {J.}~\bibnamefont {Brand}},\
  }\bibfield  {title} {\bibinfo {title} {Resonant trapping in the transport of
  a matter-wave soliton through a quantum well},\ }\href
  {https://doi.org/10.1103/PhysRevA.81.033614} {\bibfield  {journal} {\bibinfo
  {journal} {Phys. Rev. A}\ }\textbf {\bibinfo {volume} {81}},\ \bibinfo
  {pages} {033614} (\bibinfo {year} {2010})}\BibitemShut {NoStop}%
\bibitem [{\citenamefont {Damgaard~Hansen}\ \emph {et~al.}(2021)\citenamefont
  {Damgaard~Hansen}, \citenamefont {Nygaard},\ and\ \citenamefont
  {M{\o}lmer}}]{B4}%
  \BibitemOpen
  \bibfield  {author} {\bibinfo {author} {\bibfnamefont {S.}~\bibnamefont
  {Damgaard~Hansen}}, \bibinfo {author} {\bibfnamefont {N.}~\bibnamefont
  {Nygaard}},\ and\ \bibinfo {author} {\bibfnamefont {K.}~\bibnamefont
  {M{\o}lmer}},\ }\bibfield  {title} {\bibinfo {title} {Scattering of matter
  wave solitons on localized potentials},\ }\href@noop {} {\bibfield  {journal}
  {\bibinfo  {journal} {Applied Sciences}\ }\textbf {\bibinfo {volume} {11}}
  (\bibinfo {year} {2021})}\BibitemShut {NoStop}%
\bibitem [{\citenamefont {Frantzeskakis}\ \emph {et~al.}(2002)\citenamefont
  {Frantzeskakis}, \citenamefont {Theocharis}, \citenamefont {Diakonos},
  \citenamefont {Schmelcher},\ and\ \citenamefont {Kivshar}}]{B5}%
  \BibitemOpen
  \bibfield  {author} {\bibinfo {author} {\bibfnamefont {D.~J.}\ \bibnamefont
  {Frantzeskakis}}, \bibinfo {author} {\bibfnamefont {G.}~\bibnamefont
  {Theocharis}}, \bibinfo {author} {\bibfnamefont {F.~K.}\ \bibnamefont
  {Diakonos}}, \bibinfo {author} {\bibfnamefont {P.}~\bibnamefont
  {Schmelcher}},\ and\ \bibinfo {author} {\bibfnamefont {Y.~S.}\ \bibnamefont
  {Kivshar}},\ }\bibfield  {title} {\bibinfo {title} {Interaction of dark
  solitons with localized impurities in Bose-Einstein condensates},\ }\href
  {https://doi.org/10.1103/PhysRevA.66.053608} {\bibfield  {journal} {\bibinfo
  {journal} {Phys. Rev. A}\ }\textbf {\bibinfo {volume} {66}},\ \bibinfo
  {pages} {053608} (\bibinfo {year} {2002})}\BibitemShut {NoStop}%
\bibitem [{\citenamefont {Helm}\ \emph {et~al.}(2012)\citenamefont {Helm},
  \citenamefont {Billam},\ and\ \citenamefont {Gardiner}}]{A10}%
  \BibitemOpen
  \bibfield  {author} {\bibinfo {author} {\bibfnamefont {J.~L.}\ \bibnamefont
  {Helm}}, \bibinfo {author} {\bibfnamefont {T.~P.}\ \bibnamefont {Billam}},\
  and\ \bibinfo {author} {\bibfnamefont {S.~A.}\ \bibnamefont {Gardiner}},\
  }\bibfield  {title} {\bibinfo {title} {Bright matter-wave soliton collisions
  at narrow barriers},\ }\href {https://doi.org/10.1103/PhysRevA.85.053621}
  {\bibfield  {journal} {\bibinfo  {journal} {Phys. Rev. A}\ }\textbf {\bibinfo
  {volume} {85}},\ \bibinfo {pages} {053621} (\bibinfo {year}
  {2012})}\BibitemShut {NoStop}%
\bibitem [{\citenamefont {Gertjerenken}\ \emph {et~al.}(2012)\citenamefont
  {Gertjerenken}, \citenamefont {Billam}, \citenamefont {Khaykovich},\ and\
  \citenamefont {Weiss}}]{A11}%
  \BibitemOpen
  \bibfield  {author} {\bibinfo {author} {\bibfnamefont {B.}~\bibnamefont
  {Gertjerenken}}, \bibinfo {author} {\bibfnamefont {T.~P.}\ \bibnamefont
  {Billam}}, \bibinfo {author} {\bibfnamefont {L.}~\bibnamefont {Khaykovich}},\
  and\ \bibinfo {author} {\bibfnamefont {C.}~\bibnamefont {Weiss}},\ }\bibfield
   {title} {\bibinfo {title} {Scattering bright solitons: Quantum versus
  mean-field behavior},\ }\href {https://doi.org/10.1103/PhysRevA.86.033608}
  {\bibfield  {journal} {\bibinfo  {journal} {Phys. Rev. A}\ }\textbf {\bibinfo
  {volume} {86}},\ \bibinfo {pages} {033608} (\bibinfo {year}
  {2012})}\BibitemShut {NoStop}%
\bibitem [{\citenamefont {Holmer}\ \emph {et~al.}()\citenamefont {Holmer},
  \citenamefont {Marzuola},\ and\ \citenamefont {Zworski}}]{A12}%
  \BibitemOpen
  \bibfield  {author} {\bibinfo {author} {\bibfnamefont {J.}~\bibnamefont
  {Holmer}}, \bibinfo {author} {\bibfnamefont {J.}~\bibnamefont {Marzuola}},\
  and\ \bibinfo {author} {\bibfnamefont {M.}~\bibnamefont {Zworski}},\
  }\bibfield  {title} {\bibinfo {title} {Fast soliton scattering by delta
  impurities},\ }\href@noop {} {\bibfield  {journal} {\bibinfo  {journal}
  {Communications in Mathematical Physics}\ }\textbf {\bibinfo {volume}
  {274}},\ \bibinfo {pages} {187}}\BibitemShut {NoStop}%
\bibitem [{\citenamefont {Martin}\ and\ \citenamefont
  {Ruostekoski}(2012)}]{A13}%
  \BibitemOpen
  \bibfield  {author} {\bibinfo {author} {\bibfnamefont {A.~D.}\ \bibnamefont
  {Martin}}\ and\ \bibinfo {author} {\bibfnamefont {J.}~\bibnamefont
  {Ruostekoski}},\ }\bibfield  {title} {\bibinfo {title} {Quantum dynamics of
  atomic bright solitons under splitting and recollision, and implications for
  interferometry},\ }\href {https://doi.org/10.1088/1367-2630/14/4/043040}
  {\bibfield  {journal} {\bibinfo  {journal} {New Journal of Physics}\ }\textbf
  {\bibinfo {volume} {14}},\ \bibinfo {pages} {043040} (\bibinfo {year}
  {2012})}\BibitemShut {NoStop}%
\bibitem [{\citenamefont {Engels}\ and\ \citenamefont
  {Atherton}(2007{\natexlab{a}})}]{A8}%
  \BibitemOpen
  \bibfield  {author} {\bibinfo {author} {\bibfnamefont {P.}~\bibnamefont
  {Engels}}\ and\ \bibinfo {author} {\bibfnamefont {C.}~\bibnamefont
  {Atherton}},\ }\bibfield  {title} {\bibinfo {title} {Stationary and
  nonstationary fluid flow of a Bose-Einstein condensate through a penetrable
  barrier},\ }\href {https://doi.org/10.1103/PhysRevLett.99.160405} {\bibfield
  {journal} {\bibinfo  {journal} {Phys. Rev. Lett.}\ }\textbf {\bibinfo
  {volume} {99}},\ \bibinfo {pages} {160405} (\bibinfo {year}
  {2007}{\natexlab{a}})}\BibitemShut {NoStop}%
\bibitem [{\citenamefont {Dries}\ \emph
  {et~al.}(2010{\natexlab{a}})\citenamefont {Dries}, \citenamefont {Pollack},
  \citenamefont {Hitchcock},\ and\ \citenamefont {Hulet}}]{A9}%
  \BibitemOpen
  \bibfield  {author} {\bibinfo {author} {\bibfnamefont {D.}~\bibnamefont
  {Dries}}, \bibinfo {author} {\bibfnamefont {S.~E.}\ \bibnamefont {Pollack}},
  \bibinfo {author} {\bibfnamefont {J.~M.}\ \bibnamefont {Hitchcock}},\ and\
  \bibinfo {author} {\bibfnamefont {R.~G.}\ \bibnamefont {Hulet}},\ }\bibfield
  {title} {\bibinfo {title} {Dissipative transport of a Bose-Einstein
  condensate},\ }\href {https://doi.org/10.1103/PhysRevA.82.033603} {\bibfield
  {journal} {\bibinfo  {journal} {Phys. Rev. A}\ }\textbf {\bibinfo {volume}
  {82}},\ \bibinfo {pages} {033603} (\bibinfo {year}
  {2010}{\natexlab{a}})}\BibitemShut {NoStop}%
\bibitem [{\citenamefont {Weller}\ \emph {et~al.}(2008)\citenamefont {Weller},
  \citenamefont {Ronzheimer}, \citenamefont {Gross}, \citenamefont {Esteve},
  \citenamefont {Oberthaler}, \citenamefont {Frantzeskakis}, \citenamefont
  {Theocharis},\ and\ \citenamefont
  {Kevrekidis}}]{Kevrekidis-expt-oscillating-solitons}%
  \BibitemOpen
  \bibfield  {author} {\bibinfo {author} {\bibfnamefont {A.}~\bibnamefont
  {Weller}}, \bibinfo {author} {\bibfnamefont {J.~P.}\ \bibnamefont
  {Ronzheimer}}, \bibinfo {author} {\bibfnamefont {C.}~\bibnamefont {Gross}},
  \bibinfo {author} {\bibfnamefont {J.}~\bibnamefont {Esteve}}, \bibinfo
  {author} {\bibfnamefont {M.~K.}\ \bibnamefont {Oberthaler}}, \bibinfo
  {author} {\bibfnamefont {D.~J.}\ \bibnamefont {Frantzeskakis}}, \bibinfo
  {author} {\bibfnamefont {G.}~\bibnamefont {Theocharis}},\ and\ \bibinfo
  {author} {\bibfnamefont {P.~G.}\ \bibnamefont {Kevrekidis}},\ }\bibfield
  {title} {\bibinfo {title} {Experimental observation of oscillating and
  interacting matter wave dark solitons},\ }\href
  {https://doi.org/10.1103/PhysRevLett.101.130401} {\bibfield  {journal}
  {\bibinfo  {journal} {Phys. Rev. Lett.}\ }\textbf {\bibinfo {volume} {101}},\
  \bibinfo {pages} {130401} (\bibinfo {year} {2008})}\BibitemShut {NoStop}%
\bibitem [{\citenamefont {Weiss}\ and\ \citenamefont
  {Castin}(2009)}]{Castin-nonlocal-superposition}%
  \BibitemOpen
  \bibfield  {author} {\bibinfo {author} {\bibfnamefont {C.}~\bibnamefont
  {Weiss}}\ and\ \bibinfo {author} {\bibfnamefont {Y.}~\bibnamefont {Castin}},\
  }\bibfield  {title} {\bibinfo {title} {Creation and detection of a mesoscopic
  gas in a nonlocal quantum superposition},\ }\href
  {https://doi.org/10.1103/PhysRevLett.102.010403} {\bibfield  {journal}
  {\bibinfo  {journal} {Phys. Rev. Lett.}\ }\textbf {\bibinfo {volume} {102}},\
  \bibinfo {pages} {010403} (\bibinfo {year} {2009})}\BibitemShut {NoStop}%
\bibitem [{\citenamefont {Streltsov}\ \emph {et~al.}(2009)\citenamefont
  {Streltsov}, \citenamefont {Alon},\ and\ \citenamefont
  {Cederbaum}}]{Cederbaum-BEC-attractive}%
  \BibitemOpen
  \bibfield  {author} {\bibinfo {author} {\bibfnamefont {A.~I.}\ \bibnamefont
  {Streltsov}}, \bibinfo {author} {\bibfnamefont {O.~E.}\ \bibnamefont
  {Alon}},\ and\ \bibinfo {author} {\bibfnamefont {L.~S.}\ \bibnamefont
  {Cederbaum}},\ }\bibfield  {title} {\bibinfo {title} {Scattering of an
  attractive Bose Einstein condensate from a barrier: Formation of quantum
  superposition states},\ }\href {https://doi.org/10.1103/PhysRevA.80.043616}
  {\bibfield  {journal} {\bibinfo  {journal} {Phys. Rev. A}\ }\textbf {\bibinfo
  {volume} {80}},\ \bibinfo {pages} {043616} (\bibinfo {year}
  {2009})}\BibitemShut {NoStop}%
\bibitem [{\citenamefont {Luo}\ \emph {et~al.}(2020)\citenamefont {Luo},
  \citenamefont {Jin}, \citenamefont {Nguyen}, \citenamefont {Malomed},
  \citenamefont {Marchukov}, \citenamefont {Yurovsky}, \citenamefont {Dunjko},
  \citenamefont {Olshanii},\ and\ \citenamefont
  {Hulet}}]{Olashanii-Hulet-breathers}%
  \BibitemOpen
  \bibfield  {author} {\bibinfo {author} {\bibfnamefont {D.}~\bibnamefont
  {Luo}}, \bibinfo {author} {\bibfnamefont {Y.}~\bibnamefont {Jin}}, \bibinfo
  {author} {\bibfnamefont {J.~H.~V.}\ \bibnamefont {Nguyen}}, \bibinfo {author}
  {\bibfnamefont {B.~A.}\ \bibnamefont {Malomed}}, \bibinfo {author}
  {\bibfnamefont {O.~V.}\ \bibnamefont {Marchukov}}, \bibinfo {author}
  {\bibfnamefont {V.~A.}\ \bibnamefont {Yurovsky}}, \bibinfo {author}
  {\bibfnamefont {V.}~\bibnamefont {Dunjko}}, \bibinfo {author} {\bibfnamefont
  {M.}~\bibnamefont {Olshanii}},\ and\ \bibinfo {author} {\bibfnamefont
  {R.~G.}\ \bibnamefont {Hulet}},\ }\bibfield  {title} {\bibinfo {title}
  {Creation and characterization of matter-wave breathers},\ }\href
  {https://doi.org/10.1103/PhysRevLett.125.183902} {\bibfield  {journal}
  {\bibinfo  {journal} {Phys. Rev. Lett.}\ }\textbf {\bibinfo {volume} {125}},\
  \bibinfo {pages} {183902} (\bibinfo {year} {2020})}\BibitemShut {NoStop}%
\bibitem [{\citenamefont {Marchukov}\ \emph {et~al.}(2020)\citenamefont
  {Marchukov}, \citenamefont {Malomed}, \citenamefont {Dunjko}, \citenamefont
  {Ruhl}, \citenamefont {Olshanii}, \citenamefont {Hulet},\ and\ \citenamefont
  {Yurovsky}}]{Olashanii-Hulet-fluctuations}%
  \BibitemOpen
  \bibfield  {author} {\bibinfo {author} {\bibfnamefont {O.~V.}\ \bibnamefont
  {Marchukov}}, \bibinfo {author} {\bibfnamefont {B.~A.}\ \bibnamefont
  {Malomed}}, \bibinfo {author} {\bibfnamefont {V.}~\bibnamefont {Dunjko}},
  \bibinfo {author} {\bibfnamefont {J.}~\bibnamefont {Ruhl}}, \bibinfo {author}
  {\bibfnamefont {M.}~\bibnamefont {Olshanii}}, \bibinfo {author}
  {\bibfnamefont {R.~G.}\ \bibnamefont {Hulet}},\ and\ \bibinfo {author}
  {\bibfnamefont {V.~A.}\ \bibnamefont {Yurovsky}},\ }\bibfield  {title}
  {\bibinfo {title} {Quantum fluctuations of the center of mass and relative
  parameters of nonlinear Schr\"odinger breathers},\ }\href
  {https://doi.org/10.1103/PhysRevLett.125.050405} {\bibfield  {journal}
  {\bibinfo  {journal} {Phys. Rev. Lett.}\ }\textbf {\bibinfo {volume} {125}},\
  \bibinfo {pages} {050405} (\bibinfo {year} {2020})}\BibitemShut {NoStop}%
\bibitem [{\citenamefont {Mollenauer}\ \emph {et~al.}(1980)\citenamefont
  {Mollenauer}, \citenamefont {Stolen},\ and\ \citenamefont
  {Gordon}}]{Optical-breathers}%
  \BibitemOpen
  \bibfield  {author} {\bibinfo {author} {\bibfnamefont {L.~F.}\ \bibnamefont
  {Mollenauer}}, \bibinfo {author} {\bibfnamefont {R.~H.}\ \bibnamefont
  {Stolen}},\ and\ \bibinfo {author} {\bibfnamefont {J.~P.}\ \bibnamefont
  {Gordon}},\ }\bibfield  {title} {\bibinfo {title} {Experimental observation
  of picosecond pulse narrowing and solitons in optical fibers},\ }\href
  {https://doi.org/10.1103/PhysRevLett.45.1095} {\bibfield  {journal} {\bibinfo
   {journal} {Phys. Rev. Lett.}\ }\textbf {\bibinfo {volume} {45}},\ \bibinfo
  {pages} {1095} (\bibinfo {year} {1980})}\BibitemShut {NoStop}%
\bibitem [{\citenamefont {Seaman}\ \emph
  {et~al.}(2005{\natexlab{b}})\citenamefont {Seaman}, \citenamefont {Carr},\
  and\ \citenamefont {Holland}}]{A3}%
  \BibitemOpen
  \bibfield  {author} {\bibinfo {author} {\bibfnamefont {B.~T.}\ \bibnamefont
  {Seaman}}, \bibinfo {author} {\bibfnamefont {L.~D.}\ \bibnamefont {Carr}},\
  and\ \bibinfo {author} {\bibfnamefont {M.~J.}\ \bibnamefont {Holland}},\
  }\bibfield  {title} {\bibinfo {title} {Effect of a potential step or impurity
  on the Bose-Einstein condensate mean field},\ }\href
  {https://doi.org/10.1103/PhysRevA.71.033609} {\bibfield  {journal} {\bibinfo
  {journal} {Phys. Rev. A}\ }\textbf {\bibinfo {volume} {71}},\ \bibinfo
  {pages} {033609} (\bibinfo {year} {2005}{\natexlab{b}})}\BibitemShut
  {NoStop}%
\bibitem [{\citenamefont {Hakim}(1997)}]{Hakim-NLSE_delta_potential}%
  \BibitemOpen
  \bibfield  {author} {\bibinfo {author} {\bibfnamefont {V.}~\bibnamefont
  {Hakim}},\ }\bibfield  {title} {\bibinfo {title} {Nonlinear Schr\"odinger
  flow past an obstacle in one dimension},\ }\href
  {https://doi.org/10.1103/PhysRevE.55.2835} {\bibfield  {journal} {\bibinfo
  {journal} {Phys. Rev. E}\ }\textbf {\bibinfo {volume} {55}},\ \bibinfo
  {pages} {2835} (\bibinfo {year} {1997})}\BibitemShut {NoStop}%
\bibitem [{\citenamefont {Pavloff}(2002)}]{Pavlov-BEC_delta_potential}%
  \BibitemOpen
  \bibfield  {author} {\bibinfo {author} {\bibfnamefont {N.}~\bibnamefont
  {Pavloff}},\ }\bibfield  {title} {\bibinfo {title} {Breakdown of
  superfluidity of an atom laser past an obstacle},\ }\href
  {https://doi.org/10.1103/PhysRevA.66.013610} {\bibfield  {journal} {\bibinfo
  {journal} {Phys. Rev. A}\ }\textbf {\bibinfo {volume} {66}},\ \bibinfo
  {pages} {013610} (\bibinfo {year} {2002})}\BibitemShut {NoStop}%
\bibitem [{\citenamefont {Carr}\ \emph {et~al.}(2001)\citenamefont {Carr},
  \citenamefont {Mahmud},\ and\ \citenamefont {Reinhardt}}]{A5}%
  \BibitemOpen
  \bibfield  {author} {\bibinfo {author} {\bibfnamefont {L.~D.}\ \bibnamefont
  {Carr}}, \bibinfo {author} {\bibfnamefont {K.~W.}\ \bibnamefont {Mahmud}},\
  and\ \bibinfo {author} {\bibfnamefont {W.~P.}\ \bibnamefont {Reinhardt}},\
  }\bibfield  {title} {\bibinfo {title} {Tunable tunneling: An application of
  stationary states of Bose-Einstein condensates in traps of finite depth},\
  }\href {https://doi.org/10.1103/PhysRevA.64.033603} {\bibfield  {journal}
  {\bibinfo  {journal} {Phys. Rev. A}\ }\textbf {\bibinfo {volume} {64}},\
  \bibinfo {pages} {033603} (\bibinfo {year} {2001})}\BibitemShut {NoStop}%
\bibitem [{\citenamefont {Rapedius}\ \emph {et~al.}(2006)\citenamefont
  {Rapedius}, \citenamefont {Witthaut},\ and\ \citenamefont {Korsch}}]{A6}%
  \BibitemOpen
  \bibfield  {author} {\bibinfo {author} {\bibfnamefont {K.}~\bibnamefont
  {Rapedius}}, \bibinfo {author} {\bibfnamefont {D.}~\bibnamefont {Witthaut}},\
  and\ \bibinfo {author} {\bibfnamefont {H.~J.}\ \bibnamefont {Korsch}},\
  }\bibfield  {title} {\bibinfo {title} {Analytical study of resonant transport
  of Bose-Einstein condensates},\ }\href
  {https://doi.org/10.1103/PhysRevA.73.033608} {\bibfield  {journal} {\bibinfo
  {journal} {Phys. Rev. A}\ }\textbf {\bibinfo {volume} {73}},\ \bibinfo
  {pages} {033608} (\bibinfo {year} {2006})}\BibitemShut {NoStop}%
\bibitem [{\citenamefont {Ishkhanyan}\ and\ \citenamefont
  {Krainov}(2009)}]{A7}%
  \BibitemOpen
  \bibfield  {author} {\bibinfo {author} {\bibfnamefont {H.~A.}\ \bibnamefont
  {Ishkhanyan}}\ and\ \bibinfo {author} {\bibfnamefont {V.~P.}\ \bibnamefont
  {Krainov}},\ }\bibfield  {title} {\bibinfo {title} {Multiple-scale analysis
  for resonance reflection by a one-dimensional rectangular barrier in the
  Gross-Pitaevskii problem},\ }\href
  {https://doi.org/10.1103/PhysRevA.80.045601} {\bibfield  {journal} {\bibinfo
  {journal} {Phys. Rev. A}\ }\textbf {\bibinfo {volume} {80}},\ \bibinfo
  {pages} {045601} (\bibinfo {year} {2009})}\BibitemShut {NoStop}%
\bibitem [{\citenamefont {Carr}\ \emph {et~al.}(2012)\citenamefont {Carr},
  \citenamefont {Miller}, \citenamefont {Bolton},\ and\ \citenamefont
  {Strong}}]{A0}%
  \BibitemOpen
  \bibfield  {author} {\bibinfo {author} {\bibfnamefont {L.~D.}\ \bibnamefont
  {Carr}}, \bibinfo {author} {\bibfnamefont {R.~R.}\ \bibnamefont {Miller}},
  \bibinfo {author} {\bibfnamefont {D.~R.}\ \bibnamefont {Bolton}},\ and\
  \bibinfo {author} {\bibfnamefont {S.~A.}\ \bibnamefont {Strong}},\ }\bibfield
   {title} {\bibinfo {title} {Nonlinear scattering of a Bose-Einstein
  condensate on a rectangular barrier},\ }\href
  {https://doi.org/10.1103/PhysRevA.86.023621} {\bibfield  {journal} {\bibinfo
  {journal} {Phys. Rev. A}\ }\textbf {\bibinfo {volume} {86}},\ \bibinfo
  {pages} {023621} (\bibinfo {year} {2012})}\BibitemShut {NoStop}%
\bibitem [{\citenamefont {Piazza}\ \emph {et~al.}(2010)\citenamefont {Piazza},
  \citenamefont {Collins},\ and\ \citenamefont {Smerzi}}]{A15}%
  \BibitemOpen
  \bibfield  {author} {\bibinfo {author} {\bibfnamefont {F.}~\bibnamefont
  {Piazza}}, \bibinfo {author} {\bibfnamefont {L.~A.}\ \bibnamefont
  {Collins}},\ and\ \bibinfo {author} {\bibfnamefont {A.}~\bibnamefont
  {Smerzi}},\ }\bibfield  {title} {\bibinfo {title} {Current-phase relation of
  a Bose-Einstein condensate flowing through a weak link},\ }\href
  {https://doi.org/10.1103/PhysRevA.81.033613} {\bibfield  {journal} {\bibinfo
  {journal} {Phys. Rev. A}\ }\textbf {\bibinfo {volume} {81}},\ \bibinfo
  {pages} {033613} (\bibinfo {year} {2010})}\BibitemShut {NoStop}%
\bibitem [{\citenamefont {Byrd}\ and\ \citenamefont
  {Friedman}(1954)}]{Handbook_elliptic}%
  \BibitemOpen
  \bibfield  {author} {\bibinfo {author} {\bibfnamefont {P.~F.}\ \bibnamefont
  {Byrd}}\ and\ \bibinfo {author} {\bibfnamefont {M.~D.}\ \bibnamefont
  {Friedman}},\ }\href@noop {} {\emph {\bibinfo {title} {Hanbook of Elliptic
  Integrals for Engineers and Physicists}}},\ \bibinfo {edition} {1st}\ ed.\
  (\bibinfo  {publisher} {Springer-Verlag},\ \bibinfo {address} {Berlin},\
  \bibinfo {year} {1954})\BibitemShut {NoStop}%
\bibitem [{\citenamefont {Engels}\ and\ \citenamefont
  {Atherton}(2007{\natexlab{b}})}]{BEC-flow-Atherton}%
  \BibitemOpen
  \bibfield  {author} {\bibinfo {author} {\bibfnamefont {P.}~\bibnamefont
  {Engels}}\ and\ \bibinfo {author} {\bibfnamefont {C.}~\bibnamefont
  {Atherton}},\ }\bibfield  {title} {\bibinfo {title} {Stationary and
  nonstationary fluid flow of a Bose-Einstein condensate through a penetrable
  barrier},\ }\href {https://doi.org/10.1103/PhysRevLett.99.160405} {\bibfield
  {journal} {\bibinfo  {journal} {Phys. Rev. Lett.}\ }\textbf {\bibinfo
  {volume} {99}},\ \bibinfo {pages} {160405} (\bibinfo {year}
  {2007}{\natexlab{b}})}\BibitemShut {NoStop}%
\bibitem [{\citenamefont {Dries}\ \emph
  {et~al.}(2010{\natexlab{b}})\citenamefont {Dries}, \citenamefont {Pollack},
  \citenamefont {Hitchcock},\ and\ \citenamefont
  {Hulet}}]{Hulet-BEC-transport}%
  \BibitemOpen
  \bibfield  {author} {\bibinfo {author} {\bibfnamefont {D.}~\bibnamefont
  {Dries}}, \bibinfo {author} {\bibfnamefont {S.~E.}\ \bibnamefont {Pollack}},
  \bibinfo {author} {\bibfnamefont {J.~M.}\ \bibnamefont {Hitchcock}},\ and\
  \bibinfo {author} {\bibfnamefont {R.~G.}\ \bibnamefont {Hulet}},\ }\bibfield
  {title} {\bibinfo {title} {Dissipative transport of a Bose-Einstein
  condensate},\ }\href {https://doi.org/10.1103/PhysRevA.82.033603} {\bibfield
  {journal} {\bibinfo  {journal} {Phys. Rev. A}\ }\textbf {\bibinfo {volume}
  {82}},\ \bibinfo {pages} {033603} (\bibinfo {year}
  {2010}{\natexlab{b}})}\BibitemShut {NoStop}%
\bibitem [{\citenamefont {Ramanathan}\ \emph {et~al.}(2011)\citenamefont
  {Ramanathan}, \citenamefont {Wright}, \citenamefont {Muniz}, \citenamefont
  {Zelan}, \citenamefont {Hill}, \citenamefont {Lobb}, \citenamefont
  {Helmerson}, \citenamefont {Phillips},\ and\ \citenamefont
  {Campbell}}]{ramanathan}%
  \BibitemOpen
  \bibfield  {author} {\bibinfo {author} {\bibfnamefont {A.}~\bibnamefont
  {Ramanathan}}, \bibinfo {author} {\bibfnamefont {K.~C.}\ \bibnamefont
  {Wright}}, \bibinfo {author} {\bibfnamefont {S.~R.}\ \bibnamefont {Muniz}},
  \bibinfo {author} {\bibfnamefont {M.}~\bibnamefont {Zelan}}, \bibinfo
  {author} {\bibfnamefont {W.~T.}\ \bibnamefont {Hill}}, \bibinfo {author}
  {\bibfnamefont {C.~J.}\ \bibnamefont {Lobb}}, \bibinfo {author}
  {\bibfnamefont {K.}~\bibnamefont {Helmerson}}, \bibinfo {author}
  {\bibfnamefont {W.~D.}\ \bibnamefont {Phillips}},\ and\ \bibinfo {author}
  {\bibfnamefont {G.~K.}\ \bibnamefont {Campbell}},\ }\bibfield  {title}
  {\bibinfo {title} {Superflow in a toroidal Bose-Einstein condensate: An atom
  circuit with a tunable weak link},\ }\href
  {https://doi.org/10.1103/PhysRevLett.106.130401} {\bibfield  {journal}
  {\bibinfo  {journal} {Phys. Rev. Lett.}\ }\textbf {\bibinfo {volume} {106}},\
  \bibinfo {pages} {130401} (\bibinfo {year} {2011})}\BibitemShut {NoStop}%
\bibitem [{\citenamefont {Wright}\ \emph {et~al.}(2013)\citenamefont {Wright},
  \citenamefont {Blakestad}, \citenamefont {Lobb}, \citenamefont {Phillips},\
  and\ \citenamefont {Campbell}}]{Phillips_Campbell_superfluid_2013}%
  \BibitemOpen
  \bibfield  {author} {\bibinfo {author} {\bibfnamefont {K.~C.}\ \bibnamefont
  {Wright}}, \bibinfo {author} {\bibfnamefont {R.~B.}\ \bibnamefont
  {Blakestad}}, \bibinfo {author} {\bibfnamefont {C.~J.}\ \bibnamefont {Lobb}},
  \bibinfo {author} {\bibfnamefont {W.~D.}\ \bibnamefont {Phillips}},\ and\
  \bibinfo {author} {\bibfnamefont {G.~K.}\ \bibnamefont {Campbell}},\
  }\bibfield  {title} {\bibinfo {title} {Driving phase slips in a superfluid
  atom circuit with a rotating weak link},\ }\href
  {https://doi.org/10.1103/PhysRevLett.110.025302} {\bibfield  {journal}
  {\bibinfo  {journal} {Phys. Rev. Lett.}\ }\textbf {\bibinfo {volume} {110}},\
  \bibinfo {pages} {025302} (\bibinfo {year} {2013})}\BibitemShut {NoStop}%
\bibitem [{\citenamefont {Huang}\ and\ \citenamefont {Das}(2021)}]{Das-Huang}%
  \BibitemOpen
  \bibfield  {author} {\bibinfo {author} {\bibfnamefont {H.}~\bibnamefont
  {Huang}}\ and\ \bibinfo {author} {\bibfnamefont {K.~K.}\ \bibnamefont
  {Das}},\ }\bibfield  {title} {\bibinfo {title} {Effects of a rotating
  periodic lattice on coherent quantum states in a ring topology: The case of
  positive nonlinearity},\ }\href {https://doi.org/10.1103/PhysRevA.104.053320}
  {\bibfield  {journal} {\bibinfo  {journal} {Phys. Rev. A}\ }\textbf {\bibinfo
  {volume} {104}},\ \bibinfo {pages} {053320} (\bibinfo {year}
  {2021})}\BibitemShut {NoStop}%
\bibitem [{\citenamefont {Tekverk}\ \emph {et~al.}(2023)\citenamefont
  {Tekverk}, \citenamefont {Siebor}, \citenamefont {Huang},\ and\ \citenamefont
  {Das}}]{Das-Tekverk-Siebor}%
  \BibitemOpen
  \bibfield  {author} {\bibinfo {author} {\bibfnamefont {J.}~\bibnamefont
  {Tekverk}}, \bibinfo {author} {\bibfnamefont {C.}~\bibnamefont {Siebor}},
  \bibinfo {author} {\bibfnamefont {H.}~\bibnamefont {Huang}},\ and\ \bibinfo
  {author} {\bibfnamefont {K.~K.}\ \bibnamefont {Das}},\ }\bibfield  {title}
  {\bibinfo {title} {Effects of a rotating periodic lattice on coherent quantum
  states in a ring topology: The case of negative nonlinearity},\ }\href@noop
  {} {\bibfield  {journal} {\bibinfo  {journal} {(in preparation)}\ } (\bibinfo {year} {2023})}\BibitemShut {NoStop}%
\bibitem [{\citenamefont {Das}(2011)}]{Das-PRA-R-Shift}%
  \BibitemOpen
  \bibfield  {author} {\bibinfo {author} {\bibfnamefont {K.~K.}\ \bibnamefont
  {Das}},\ }\bibfield  {title} {\bibinfo {title} {Mesoscopic transport and
  interferometry with wave packets of ultracold atoms: Effects of quantum
  coherence and interactions},\ }\href
  {https://doi.org/10.1103/PhysRevA.84.031601} {\bibfield  {journal} {\bibinfo
  {journal} {Phys. Rev. A}\ }\textbf {\bibinfo {volume} {84}},\ \bibinfo
  {pages} {031601(R)} (\bibinfo {year} {2011})}\BibitemShut {NoStop}%
\bibitem [{\citenamefont {Zhang}\ and\ \citenamefont {Li}(2015)}]{A14}%
  \BibitemOpen
  \bibfield  {author} {\bibinfo {author} {\bibfnamefont {X.-R.}\ \bibnamefont
  {Zhang}}\ and\ \bibinfo {author} {\bibfnamefont {W.-D.}\ \bibnamefont {Li}},\
  }\bibfield  {title} {\bibinfo {title} {Nonlinear tunneling through a strong
  rectangular barrier},\ }\href {https://doi.org/10.1088/1674-1056/24/7/070311}
  {\bibfield  {journal} {\bibinfo  {journal} {Chinese Physics B}\ }\textbf
  {\bibinfo {volume} {24}},\ \bibinfo {pages} {070311} (\bibinfo {year}
  {2015})}\BibitemShut {NoStop}%
\bibitem [{\citenamefont {Das}\ and\ \citenamefont
  {Aubin}(2009)}]{Das-Aubin-PRL2009}%
  \BibitemOpen
  \bibfield  {author} {\bibinfo {author} {\bibfnamefont {K.~K.}\ \bibnamefont
  {Das}}\ and\ \bibinfo {author} {\bibfnamefont {S.}~\bibnamefont {Aubin}},\
  }\bibfield  {title} {\bibinfo {title} {Quantum pumping with ultracold atoms
  on microchips: Fermions versus bosons},\ }\href
  {https://doi.org/10.1103/PhysRevLett.103.123007} {\bibfield  {journal}
  {\bibinfo  {journal} {Phys. Rev. Lett.}\ }\textbf {\bibinfo {volume} {103}},\
  \bibinfo {pages} {123007} (\bibinfo {year} {2009})}\BibitemShut {NoStop}%
\end{thebibliography}

%

\end{document}